BTU Cottbus–Senftenberg
Faculty 1 – Institute for Computer Science

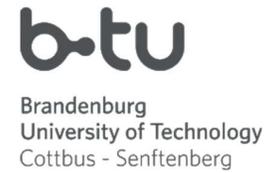

– Master Thesis –

# Security Issues on the OpenPLC project and corresponding solutions

Chaerin Kim
Cyber Security

*November 18, 2023*

**Reviewer 1:** Prof. Peter Langendörfer
**Reviewer 2:** Hon. Prof. Z. Dyka
**Supervisor:** Wael Alsabbagh

# Eidesstattliche Erklärung

Der Verfasser erklärt an Eides statt, dass er die vorliegende Arbeit selbständig, ohne fremde Hilfe und ohne Benutzung anderer als der angegebenen Hilfsmittel angefertigt hat. Die aus fremden Quellen (einschließlich elektronischer Quellen) direkt oder indirekt übernommenen Gedanken sind ausnahmslos als solche kenntlich gemacht. Die Arbeit ist in gleicher oder ähnlicher Form oder auszugsweise im Rahmen einer anderen Prüfung noch nicht vorgelegt worden.

Ort, Datum                                                                 Unterschrift des Verfassers

# Acknowledgement

I would like to express my deepest gratitude to Prof. Dr. Peter Langendörfer, Hon. Prof. Dr. Zoya Dyka, and more especially to Dr. Wael Alsabbagh for the opportunity and confidence he gave me to work under his supervision. Dr. Wael Alsabbagh inspired me on the topic and encouraged me to be professional. His constructive guidance, insightful feedback, and committed support were immeasurable and appreciated. I am truly fortunate to have had the privilege of learning under his mentorship, and his encouragement has been a driving force behind the successful completion of this work. I wish to convey my gratitude to Dr. Alsabbagh for his pivotal role in this academic journey. I also could not have undertaken this journey without my defense committee, who generously provided knowledge and expertise.

I would like to extend my gratitude to my friends and colleagues both at BTU and within IHP: Innovations for High Performance Microelectronics, appreciating the advice and comments you provided in one way or the other. Once again, I express my heartfelt thanks to each of you.

Finally, I am also thankful to my family and friends, who are the most supportive in my life. This accomplishment would not be possible without their support.

# Contents

# Abstract

Open-source Programmable Logic Controller (OpenPLC) software is designed to be vendor-neutral and run on almost any computer or low-cost embedded devices e.g., Raspberry Pi, Arduino, and other controllers. The project aims to introduce an affordable and practical alternative solution for the high cost of real hardware PLCs. It has successfully gained substantial interest within both the research and industrial communities. Due to the growth of its popularity, understanding its security vulnerabilities and implementing effective mitigation strategies becomes crucial. Through a combination of threat modeling, vulnerability analysis, and practical experiments, this thesis provides valuable insights for developers, researchers, and engineers aiming to deploy OpenPLC securely in industrial environments. To this end, this work first conducts an in-depth analysis aimed to shed light on various security challenges and vulnerabilities within the OpenPLC project. It encompasses issues such as unauthorized access, weak communication protocols, concerns regarding data integrity, the absence of robust encryption mechanisms, etc. After that, it shows the consequences of exploiting those vulnerabilities to the research community. To this end, an advanced control logic injection attack was performed. This attack maliciously modifies the user program run on the OpenPLC Runtime. The presented injection is stealthy and not detected by the legitimate user. Finally, the work introduces a security-enhanced OpenPLC software called OpenPLC Aqua. The new software is equipped with a set of security solutions designed specifically to address the vulnerabilities to which current OpenPLC versions are prone. All the attack codes and OpenPLC Aqua software are publically available for both engineers and research communities.

# List of Figures

# List of Tables

# 1  Introduction

While industrial automation is drastically growing, Programmable Logic Controller (PLC) have contributed significantly. Since most industrial processes are carried out with the help of PLCs, PLCs are considered an essential ingredient of Industrial Control System (ICS). They are used widely across various industries because they can respond instantaneously and provide simple operation and flexible programming. They are primarily deployed for automating processes and controlling critical infrastructure equipment such as motors, valves, and sensors in industries, e.g., water, energy, and transportation. Industrial facilities have a demand to increase their connectivity for performance and flexibility in the recent changing technological landscape, though they were designed for long-term use in closed environments. As a result, they are not durable against current cyber-attacks.

The most notable attack on ICS in history is Stuxnet discovered in 2010, specifically aimed at nuclear facilities in Iran [6]. It has destroyed almost 1,000 uranium centrifuges by exploiting the vulnerabilities of PLCs. The attacker has designed the program for a particular brand of PLC utilized while separating nuclear material. It is the first known malware that can destroy hardware. Stuxnet is delivered via a USB stick and finds its target that utilizes the SIEMENS program for the automation process by traveling the network. Once the target is discovered, it sends the attack command to the PLCs, inducing their self-destruction while transmitting a false feedback message to the main controller for monitoring. After Stuxnet, many attack groups exploited the modified Stuxnet to give effectual damage to critical infrastructures.

In 2013, a German steel processing factory suffered massive damage from a targeted cyber-attack [7]. For the attack, email for spear phishing and social engineering are combined to gain access to the inner system of the facilities. In this case, the adversary fulfilled the knowledge of IT security and specialized software utilized in the factory. Consequently, he succeeded in tampering with the control of the production.

Another example is named Industroyer, which paralyzed the electric grids in Ukraine [8]. It sophisticatedly targeted a specific type of substation and furnished several tactics to break the operation. In 2016, it succeeded in an outage of the electric grids for an hour. Industroyer currently came back with version 2, improving itself by customizing for each victim.

As PLCs become an attractive target of attacks, the overall security level of critical infrastructures needs to be improved because it gets threatened. However, there are several challenges for improvement. The most critical issue is that PLCs are too expensive to deploy for academic research. The high cost of PLCs discourages academic research from proceeding to advanced areas. Another problem is that they are hard to patch for following new security standards because of their design features. Patching for the existing PLCs requires a very exquisite test before it starts. However, the test for patching is challenging since PLCs are not affordable, and the patching work should not interrupt their operation.

Usually, security solutions for ICSs and IT systems have differences. Not like legacy IT systems, ICS has various restrictions to secure itself, and the goal of security is different as well [9]. IT systems prioritize to protect their data from attacks. However, ICSs, including critical infrastructures, must safeguard human safety, and it should be the first goal of their security because any accident can be linked to the destruction of human life. Since their goals are not the same, the order of their security requirements in each system is also different. Traditionally, IT systems have requirements of confidentiality, integrity, and availability in order. According to [9], not like IT systems, availability is the most significant requirement for ICSs. Therefore, transplanting security solutions designed for legacy IT systems is not an appropriate resolution for ICSs. Instead, ICS demands its own security resolutions that are tailored to them.

In order to correspond to contemporary cyber threats in industrial environments, we need an additional number of skillful engineers who are sufficiently trained and comprehend the automation process in industrial environments. However, due to their expensive cost, PLCs can still be a financial challenge for academic researchers to use as scientific testbeds, though they rise as an intriguing subject of study. It also requires time for researchers and engineers to be skillful in PLC programming and its maintenance. In 2016, Rodríguez et al. presented the result of their laboratory practices that virtualization of the Supervisory Control and Data Acquisition (SCADA) system without an actual factory could gain success for teaching students [10]. Like the former experiment, virtualizing the environment for PLCs experiment can foster more experts without financial challenges.

Another challenging part for PLC programmers is usually the integration of PLC programs for further improvement. Before implementing a new strategy on a device, they should make it as complete as possible so as not to interrupt the operation of PLCs. Otherwise, it will cost more. Meaning that, the programmers have a limited number of chances to test their strategies and find mistakes.

In recent years, therefore, the simulators or emulators of PLCs have been introduced to solve those challenges and support more affordable opportunities for the test, e.g., Do-more Designer[1], WPLSoft[2], WTE PLC Simulator and Configuration Tool[3], etc. Among them, OpenPLC is one of the most successful and outstanding simulators. It is not limited to creating and experiencing the programming but is also feasible to establish a small ICS environment. The significant advantages of the OpenPLC are that it can be run on low-cost hardware such as Raspberry Pi and provides individuals an opportunity to experience PLC programming without paying expensive costs. Subsequently, the accessibility contributes to expanding the potential user base of PLC programming. All these tools allow users to enhance their understanding of PLCs and creativity by customizing PLC programs. Additionally, they can be deployed to test control strategies in a virtual environment before implementation on real devices. By providing more opportunities for testing, the PLC programmers can prevent mistakes and reduce downtime and costs. Nonetheless, it is considered critical against various breaches, such as injection attacks, buffer overflows, unauthorized access, insecure communication channels, and more. The latent security issues pose the risk of exploitation by adversaries with malicious intentions, affecting the physical process controlled by the project in the end.

In this thesis, we highlight the disclosed vulnerabilities by a thorough investigation. The scrutinized examination reveals existing security flaws that software allows unauthorized access, modification, and control of physical processes. In addition, adversaries can access sensitive data saved in the database without an appropriate verification process because of the lack of access restrictions on the software, particularly on the *Webserver*, where the software keeps the core executable files, including the uploaded programs. By exploiting revealed vulnerabilities, attackers can perform a severe attack without exposing them to legitimate users or deceive users with fake information.

Besides, we demonstrate the design deficiency that lacks confidentiality and integrity in the communication between the OpenPLC Runtime application and users. Along with the absence of encryption for saved data, the software deploys a communication channel without encryption, not ensuring the integrity of transmitted data and being vulnerable to interception attacks.

---

[1] https://www.automationdirect.com/do-more/brx/software
[2] https://downloadcenter.deltaww.com/en-US/DownloadCenter
[3] https://www.wte.co.nz/plc-simulator.html

By exploiting discovered security issues, we perform a sophisticated control logic attack. The attack proves the fact that the software enables the access of unauthorized adversaries, incurring damage to the operation of the software. Notably, the attack does not require a root privilege and is not detected by legitimate users' eyes.

In order to mitigate divulged vulnerabilities, we introduce OpenPLC Aqua, which integrates security functions into the original version. In improving the overall security level of the project, the new software resolves the susceptibilities in the original version. It enhances the confidentiality of sensitive data using the Advanced Encryption Standard (AES) algorithm. The encrypted data is encoded into American Standard Code for Information Interchange (ASCII) letters to be compatible with older designs. The reinforced confidentiality prevents the revealing of critical data by attackers.

By performing strict access restrictions, the latest software impedes the attempt of attackers' access from external environments. Accordingly, the control policy is adopted for both Webserver and project files. Moreover, OpenPLC Aqua offers a whitelisting function to block the access of untrusted users, thereby blocking malicious activities proactively despite adversaries obtaining the registered user accounts.

OpenPLC Aqua deploys not only the encryption of saved data in the Webserver but also the encryption of the communication channel, adding the Secure Sockets Layer/Transport Layer Security (SSL/TLS) layer. It provides confidentiality and integrity of the channel while preventing sniffing attacks and eavesdropping. The presented control logic attack and OpenPLC Aqua are publically available for scientific researchers and industrial engineers.

**1.1  Problem Statement.** OpenPLC succeeded in offering an affordable and full-featured standardized alternative for scientific researchers and engineers. By providing a virtualized testbed, which is a novel solution, OpenPLC has contributed to enhancing our understanding of PLCs. Despite the great success achieved by the project, it has doubts that it is susceptible to cyberattacks on PLCs, thereby failing to establish robust security mechanisms. In order to promote academic research to gain more advanced results, it is urgent to check the susceptibility of the project. Accordingly, the OpenPLC must proceed with a meticulous investigation to disclose its vulnerabilities.

**1.2  Research Objectives.** The main objective of this thesis is to investigate the security of the OpenPLC project and then introduce a more security-enhance software that is more resistant to a wide range of cyber-attacks. To this end, we address the following research questions that reflect the results of our work.

a) Is the OpenPLC software vulnerable and prone to cyber-attacks? To address this question, we provide a comprehensive investigation, revealing several vulnerabilities within the design of the OpenPLC project. (addressed in Chapter 3).

b) Can adversaries conduct a sophisticated control logic injection attack against the OpenPLC? In this thesis, we present the first control logic injection attack that is tailored for OpenPLC and related systems. Our injection approach maliciously alters the ongoing user program, thereby causing confusion in the physical process controlled by the compromised OpenPLC. (addressed in Chapter 4).

c) Is it possible to enhance the security of the current OpenPLC software? Upon our findings in Chapter 3, we enhanced the OpenPLC software by introducing a more secure version known as OpenPLC Aqua. This new software addresses the vulnerabilities present in current OpenPLC versions while remaining compatible with the same hardware devices as the older versions. (addressed in Chapter 6)

### 1.3 Research Methodology.

a) **Atacker model** We assumed that the attacker can access to the network where the OpenPLC is located. The major tactics that used in our attack is based on help of *MITRE ATT&CK*[4]. Fig 1.1 depicts the attacker model used in the thesis.

b) **Attack Techniques** *ATT&CK* is the framework, which is archived by the organization *MITRE*, to document the possible tactics, techniques, and procedures of adversaries. The framework concentrated to address four issues; a behavior of adversaries, a suggestion of lifecycle models that fits to new types of sensors, applicability to real environments, and common taxonomy. ATT&CK is used to emulate new threats in cyber world.

**T1555 Credentials from Password Stores** Passwords are stored in several places in the location and the attacker can search the location of the password storage.

**T1040 Network Sniffing** The attacker can sniff the network traffic to gain the information of target including the credentials for authentication.

**T0812 Default Credentials** The attacker can utilize the default account offered from the service provider to control the target system. The default account may have the administrative permission.

---
[4]https://attack.mitre.org/

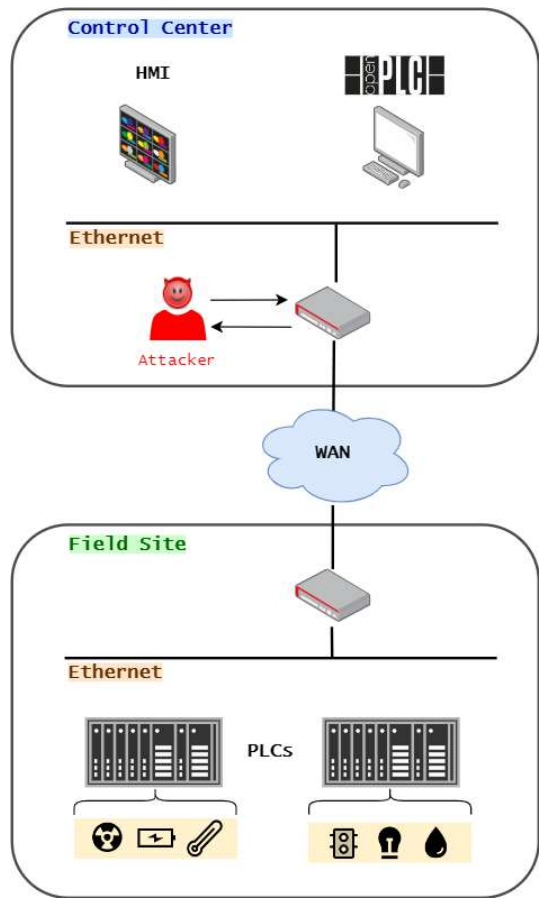

Figure 1.1: Attacker model [1]

**T1110.002 Brute Force: Password Cracking** The attacker can use password cracking technique to obtain the password in readable format.

**T1557 Adversary-in-the-Middle** The attacker can attempt to place themselves between hardware controller and PLC by abusing features of common network protocols (e.g., ARP, Domain Name System (DNS), etc.).

**T0831 Manipulation of Control** In the industrial environment, the attacker can manipulate the control of the physical process, such as set point values, tags, etc.

**T0821 Modify Controller Tasking** The attacker can interrupt the tasking of controller or the execution of their program. He can manipulate the flow of program execution.

**T0889 Modify Program** The attacker can add or update the program on PLC to affect how it interacts with physical process, peripheral devices, or other hosts on the network.

**T0845 Program Upload** The attacker can upload the program to acquire the information of industrial process.

**1.4 Contributions.** In this thesis, we take the attack approach and countermeasures presented in our former publications. Our main contributions to this thesis are summarized as follows:

a) **Dissecting the architecture of OpenPLC**: The thesis provides a comprehensive investigation of OpenPLC, revealing multiple vulnerabilities derived from the design flaws of the project.

b) **Reclaiming control logic attack approach based on OpenPLC systems**: A new control logic attack is introduced, which alters the existing user programs, causing operational error [1].

c) **Concealing the attack approach**: By making the attack stealthier, the attack becomes much more effective and powerful. Multiple security flaws of the project are exploited, thereby hiding the attack from legitimate users.

d) **Demonstrating the enhanced software, OpenPLC Aqua**: An improved software is introduced based on our discovered vulnerabilities in the project. OpenPLC Aqua, the new software, covers security issues that were found in the original project while being compatible with the same hardware devices as the current version [4].

**1.5 Scientific Publications.**

1) Alsabbagh, Wael, Kim, Chaerin and Langendörfer, Peter. Good Night, and Good Luck: A Control Logic Injection Attack on OpenPLC. published at IECON 2023- 49th Annual Conference of the IEEE Industrial Electronics Society, Singapore, Singapore, 2023, pp. 1-8, doi: 10.1109/IECON51785.2023.10312570.

2) Alsabbagh, Wael, Kim, Chaerin and Langendörfer, Peter. No Attacks Are Available: Securing the OpenPLC and Related Systems. presented at 8th IACS WS'23: 8th GI/ACM Workshop on Industrial Automation and Control Systems IACS 2023, Berlin, doi: 10.13140/RG.2.2.24570.47043.

3) Alsabbagh, Wael, Kim, Chaerin and Langendörfer, Peter. Investigating the Security of OpenPLC: Vulnerabilities, Attacks, and Mitigation Solutions. submitted at IEEE Access, 2023. doi: 10.13140/RG.2.2.34456.98566/2.

**1.6 Organization.** The rest of the thesis is structured as follows: Chapter 2 describes the architecture of OpenPLC and precedent studies that were published previously, followed by the associated security issues of OpenPLC in Chapter 3. Chapter 4 elaborates on our approach to attack the OpenPLC by exploiting the addressed vulnerabilities. In Chapter 5, the suggestion of security solutions are proposed for the identified weaknesses and attempted attacks. The experimental result after adopting our solutions is presented in Chapter 6. Finally, the thesis concludes with reflection on the insights gained from working on this thesis.

## 2  Literature Review

This chapter introduces the architecture of OpenPLC project and precedent studies related to the security of the project.

### 2.1  Background

The OpenPLC[1] is an open-source PLC emulator that allows users to create custom ICS. The project was published in 2014 [11], and now version 3 is available. The OpenPLC is designed to run on various platforms and supports different communication protocols, such as Modbus and Distributed Network Protocol (DNP3). Both are popular protocols that are used for SCADA communication.

The OpenPLC consists of three components; Editor, Runtime, and Builder. Fig 2.1 provides a high-level view of the architecture of the project. The OpenPLC *Editor* offers an alternative approach to PLC programming by allowing users to create programs using graphical and textual programming languages identified by International Electrotechnical Commission (IEC) 61131-3 standard. IEC 61131 is an IEC standard for programmable controllers published in 1993 [12]. It consisted of 10 parts to depict the standard, and Part 3 introduced basic software architecture and programming languages of PLC. The OpenPLC Editor allows five programming languages, i.e., Function Block Diagram (FBD), Ladder Diagram (LD), Structured Text (ST), Instruction List (IL), Sequential Function Chart (SFC).

Whereas the OpenPLC Editor supports all five languages to write a program, it translates the written program into an ST formatted program with the help of the PLCopen Editor. Structured Text is a textual programming language that resembles Pascal, allowing the complex structure of a program [13]. It is based on block structure and compatible with other IEC 61131-3 languages. Though ST is versatile, it is not as intuitive as other languages because it is textual.

---

[1]https://openplcproject.com/

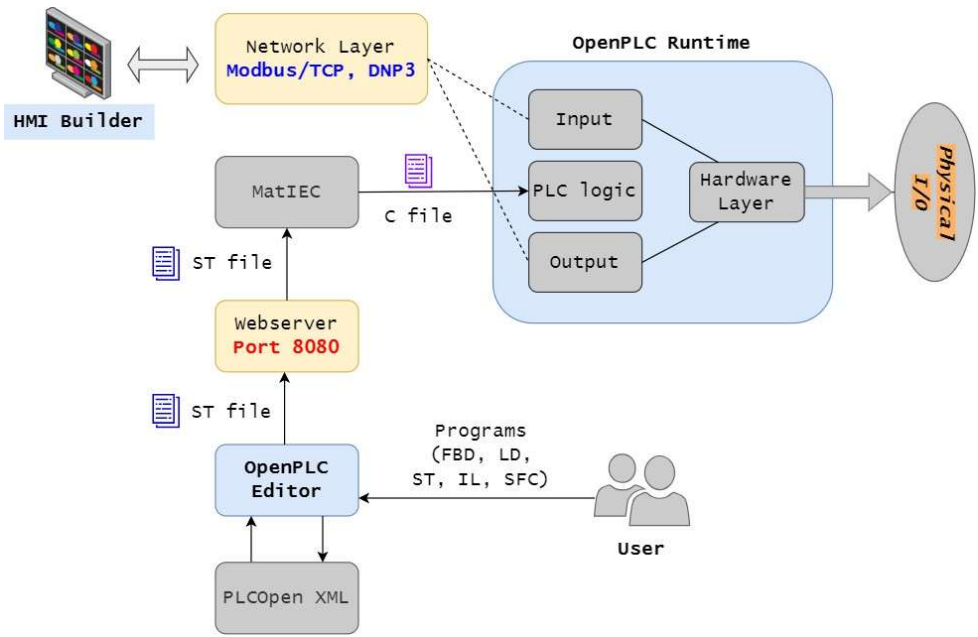

Figure 2.1: Structure of OpenPLC [1]

PLCopen Extensible Markup Language (XML)[2] allows a program to be compatible with different environments and it is particularly designed for IEC 61131-3. In order to prevent confusion, we specify that PLCopen[3] is a distinct organization where offering standards for efficiency in industrial automation. When a control logic program is written in ST format, it is copied and stored in the *Webserver*. Then, the ST file is compiled into a C file by using built-in MatIEC module.

MatIEC[4] is an open source IEC 61131-3 compiler. It is originally developed for the project named MatPLC [14]. The MatPLC project[5] was also aimed to implement the open-source PLC and to imitate the operation of PLCs. The project deployed a modular structure to emulate the performance of PLCs [15]. One of the modules is the IEC 61131-3 compiler. It intended to allow various types of PLC programming languages to interoperate and expand to C. Hence, the MatIEC compiler translates the file written in three types of PLC programming languages, i.e., ST, SFC, and IL, into ANSI C. Then, the Runtime application executes the compiled C file. While the program is running, the administrators can monitor the execution

---

[2]https://plcopen.org/technical-activities/xml-exchange
[3]https://plcopen.org/what-plcopen
[4]https://github.com/nucleron/matiec.git
[5]https://mat.sourceforge.net/

of control programs in the Human Machine Interface (HMI) builder, e.g., ScadaBR (Fig 2.2).

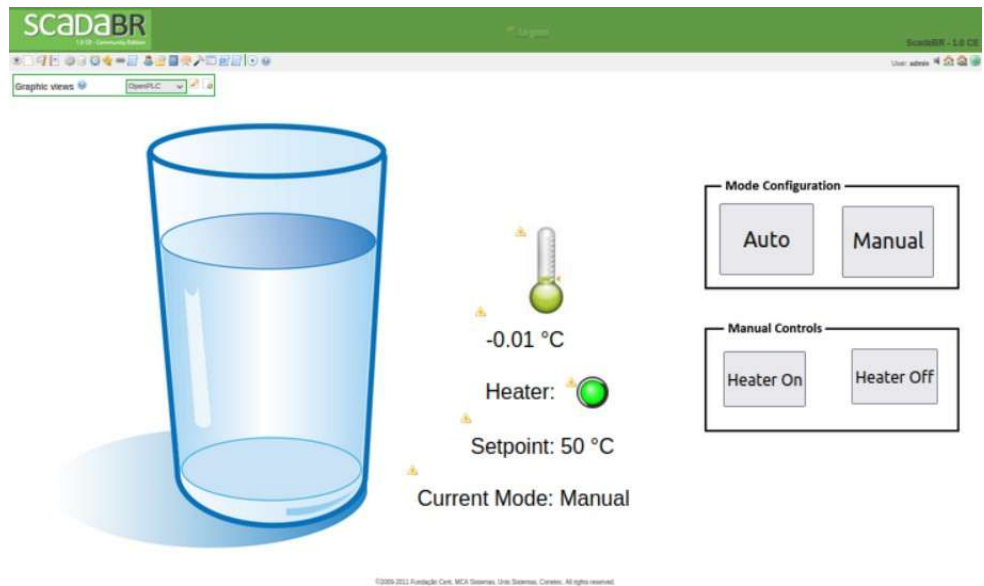

Figure 2.2: OpenPLC Builder: ScadaBR

The software utilizes Modbus/Transmission Control Protocol (TCP) or DNP3 protocol to establish a communication with SCADA software. Modbus is a data communication protocol developed for industrial applications in 1979 [16]. It is popular to deploy because it is relatively easy to use and maintain and has fewer limitations for the format of data transmission. The protocol is commonly used to connect the supervisory computer to the SCADA system. It supports four types of objects: coils, discrete inputs, input registers, and holding registers. As time passed after development, however, a problem was discovered that it could not support newly appeared data types, e.g., large binary data. Another consequential problem is that the protocol offers no security mechanisms against breaches. Another representative protocol, DNP3, is a set of communications protocols commonly deployed in electric and water supply facilities [17]. It possesses a crucial role in SCADA communication, connecting the HMI station to field devices.

ScadaBR[6] is an open-source SCADA software that can be used as an HMI monitor. It is feasible to access even from a mobile environment because it offers web interfaces to developers. The software allows control of devices, graphical monitoring, statistics, configuration of protocol, and alarms in real-time. The memory addresses of PLC inputs and outputs bind to the corresponding Modbus input and output addresses.

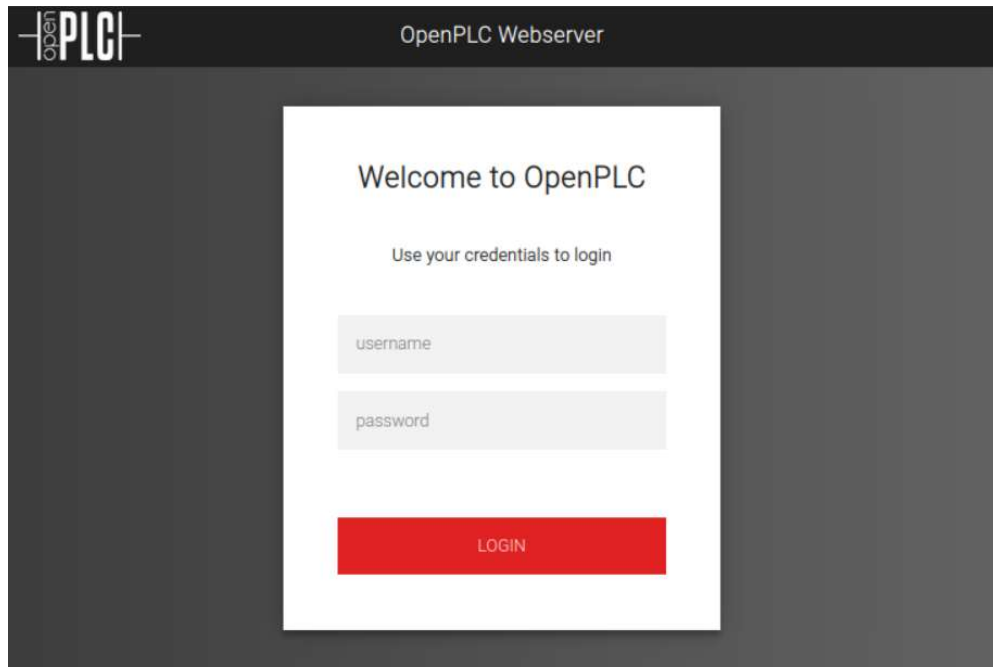

Figure 2.3: Login page

The operator can control their control logic program on the OpenPLC Runtime application, and the control of the program encompasses uploading, updating, compiling, executing, and removing. As a protective mechanism, the software verifies the authentication whenever a user accesses the application. For the first login, the project has the default account with the same username and password, "openplc". After the first login, the user can make a choice to change it or register a new account.

The *Webserver* runs on port 8080 and since it is a part of the Runtime application, the user can access the Runtime application by opening 'locahost:8080' on any web browser, e.g., Google Chrome, Firefox, Safari, etc. One of the main functions of the Webserver is to store user credentials in the database named *openplc.db*. The *openplc.db* stores not only username and password, but also all information on uploaded programs, registered devices, and custom configurations. The user credentials are stored with an entered email address, the location of the profile picture, and the unique identifier *'User_ID'*. The default account has '10' as the identifier, and 'User_ID' is used as the index to call the record of the account information. In the same manner, the uploaded program has its own *'Prog_ID'*, and the general information of the uploaded program, i.e., the title of the program, the location of copy, and the uploaded date of the program, is loaded by indexing identifier.

---

[6]http://www.scadabr.com.br/

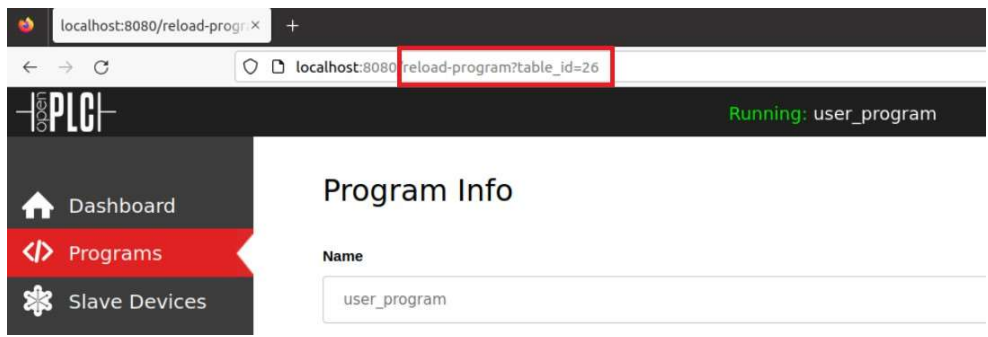

Figure 2.4: Program information page

When the application calls the information page of the uploaded program, it uses 'Prog_ID' as 'table_id' on the web address, e.g., localhost:8080/reload-program?table_id=Prog_ID (Fig 2.4). 'User_ID' is used in the same way when the information of the enrolled account is called. The more files are uploaded, the unique identification number automatically increases. The number of identifier does not decrease unless deliberately modified.

The Webserver stores executable files which operate as the main function of the project, such as scripts to compile and remove programs. The Webserver also saves the copies of uploaded ST files, etc. On the other hand, the file named 'active_program' to indicate the compiled copy is stored as well, and it is used to launch the application. If it is empty or the matching copy does not exist in the Webserver, the application cannot be launched. It is not strange to call Webserver as the nucleus of the project.

Meanwhile, the communication between user and OpenPLC Runtime application always uses Hypertext Transfer Protocol (HTTP) packets. Fig 2.5 shows the upload process in OpenPLC. When the user uploads his program file in ST format, he sends HTTP "Upload-program" packet. Then, the OpenPLC saves the copied target program with randomly generated number as title of the copy in the Webserver and responds back with "OK 200" packet. In most cases, the "OK" packet includes the HTTP status code 200. The "Upload-program-action" packet is automatically sent by user side after copying process. If the OpenPLC receives the packet, it stores the information of the program in openplc.db. The OpenPLC sends "OK" packet again if there is no problem while generating the new record in the database. Then, the compilation process is started by sending "compile-program" packet to OpenPLC. Once the compiling is started, "compilation-logs" packets are automatically transmitted several times at regular intervals until the compiling process is complete. After user gets the message that the process is finished, he selects to go back to Dashboard in manual. To run the uploaded program, the compiling process should be executed before the program is started and the user can manually starts the execution. When the "start-plc" packet is transmitted to the OpenPLC, the software responds with "OK 302" packet and redirects to the Dashboard automatically.

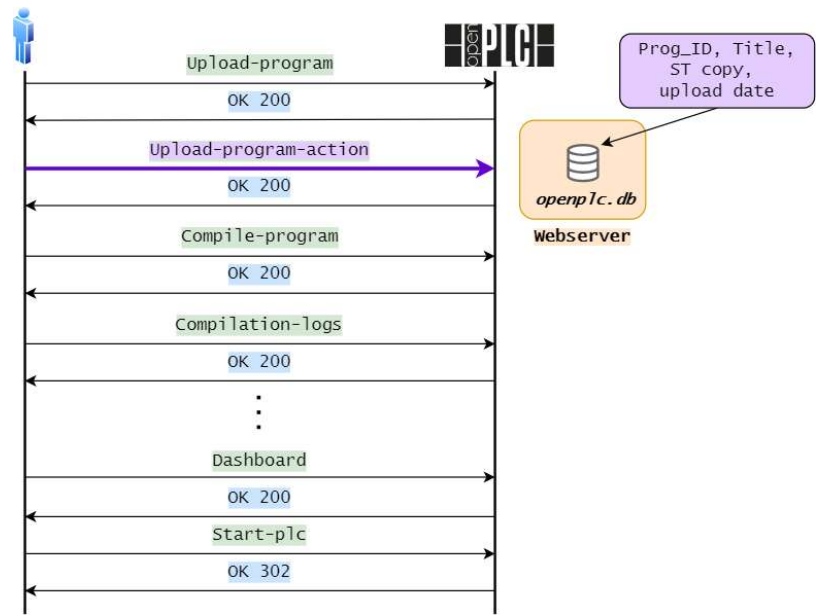

Figure 2.5: Upload sequence [1]

In conclusion, OpenPLC is an effective tool for simulating and learning industrial programming and systems without a large budget, and it is simple to use as well, providing compatibility with various web browsers. Nonetheless, it is not suitable for real-time projects with environmental or safety concerns. If real-time behavior is required for the project, the founder recommended to customize the software for its demands.

## 2.2 Related Studies

After the publication of the project, some researchers conducted the investigation about its possibility of cyber attacks, but mostly focused on the communication between the Runtime application and Builder software.

The founder of OpenPLC introduced the OpenPLC Neo to protect Modbus/TCP messages by encrypting them with AES algorithm [2]. He suggested the Neo version to expand original OpenPLC to industrial level and support wide range of PLCs. The Modbus/TCP connection between OpenPLC and external network is encrypted by AES-256 algorithm. The AES-256 encryption using in Neo version includes Secure Hash Algorithm (SHA-256) encryption for its keyphrase. Once the keyword is provided by user, SHA-256 encryption fixes the length of keyphrase into 256-bit hash, and the generated hash is used as a key of AES-256 encryption. The result of AES algorithm combines with the size of message and its initial vector that is used to the first round of encryption process. However, if the opposite party (i.e., external network) cannot decrypt the encryptped messsage from the OpenPLC, they cannot establish a communication. The Neo version configures the OpenPLC Localhost Gateway to solve the problem. It is placed in the middle of the OpenPLC Neo and the legacy SCADA software and helps SCADA software to decrypt the message. It allows to be compatible with SCADA software while covering the weak nature of Modbus/TCP protocol. Fig 2.6 offers the general view of the structure of OpenPLC Neo with features highlighted.

A few years later, he suggested the more secure version of OpenPLC Neo embedding Intrusion Prevention System (IPS) with machine learning technique [18]. He proposed placing the AES-256 encryption and IPS layers between OpenPLC and external networks (e.g., PLCs, legacy SCADA software). The same ideas of the encryption algorithm and secure gateway system are brought from [2]. After the messages go out to the external network, they are encrypted by the AES-256 algorithm, and the resulting ciphertext meets the IPS layer. The IPS layer utilizes machine learning techniques to detect suspicious packets between OpenPLC and devices. It requires the information for its performance, such as packet latency and packet processing information, which are only available at the TCP header. The trained IPS layer forwards the encrypted messages if they are valid. The author conducted three major attacks on the ICS network (interception, injection, and Denial of Service attacks) to show they are not effective on their secure version of OpenPLC anymore, and its real-time response is even better than usual commercial PLCs though it uses machine learning techniques. However, these two suggestions have a weaknesses because of its security on the symmetric encryption algorithm, so overall security level is up to the management of the key.

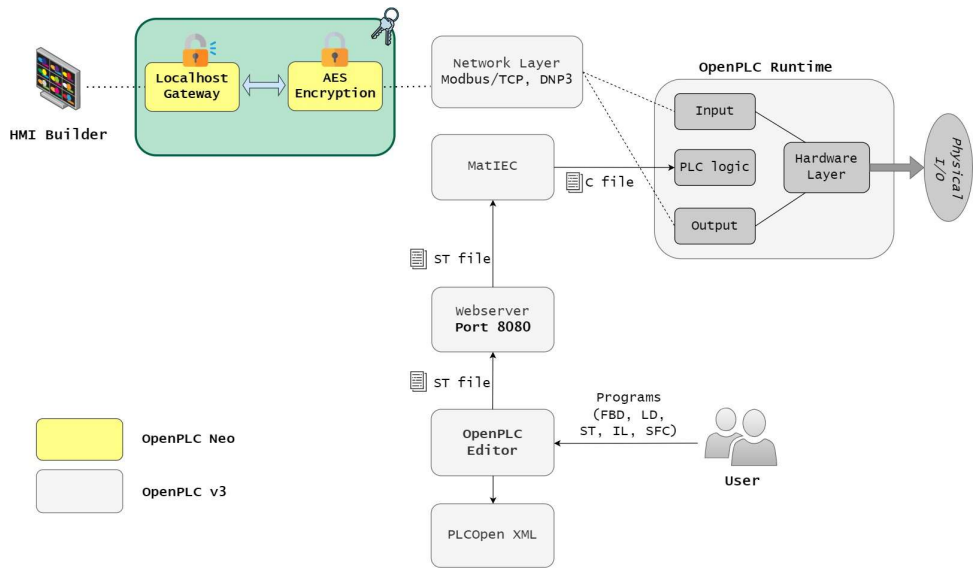

Figure 2.6: Structure of OpenPLC Neo [2]

Roomi et al. suggested the extended version of OpenPLC which supports the advanced ICS by applying IEC 61850 on the OpenPLC [19]. IEC 61850 is an international standard designed for communication protocols using in smart grid systems [20]. The purpose to expand the OpenPLC project is to get the idealized testbed for modernized smart substations. The OpenPLC only deploys traditional ICS protocols, i.e., Modbus/TCP and DNP3, as communication protocols for devices, while the author's version extends the software to adopt the advanced protocol such as Manufacturing Message Specification (MMS) for power grids.

In the following work, he presented cyber attacks on OpenPLC61850. In [21], False Data Injection (FDI) attack and a False Command Injection (FCI) attack utilizing MMS messages are presented, and both experiments were successful. Additionally, he tested the FCI attack using Modbus protocol, and his scenarios gained success in controlling the communication between OpenPLC61850 and SCADA software with dummy messages. However, he did not suggest corresponding security resolutions.

Alsabbagh et al. introduced the Modbus FCI attack by deploying the OpenPLC as a testbed

[22]. They exploited the features of Modbus lacking securing functions and executed the Man-in-the-Middle (MitM) attack to damage the SCADA communication between HMI and PLCs. Before the implementation, the adversary collected the pair of Modbus requests and responses for a long time and scanned the opened ports of each device. A more prosperous collection enabled a modified attack because each pair of Modbus packets shares the same Transaction ID. Moreover, the attacker executed Address Resolution Protocol (ARP) spoofing to deceive HMI and PLCs that the attacker's machine was their partner device for communication. When the HMI sent the Modbus request to the PLC, the adversary discarded it and transmitted his malicious request to the PLC. In the same manner, the attacker dropped the response from the PLC and sent the malicious response to the HMI instead. In addition, the attacker maintained the TCP real-time sessions and avoided TCP timeouts to keep the attack stealthy. However, the PLCs will report the anomaly if the attacker stops ARP spoofing. The authors suggested replacing the Modbus protocol with a more secure version, Modbus TLS, to mitigate the risks.

One of the most impressive works was the remote code execution attack published in [23]. The author remotely connected to the machine that installed the OpenPLC Runtime and injected the falsified command into the OpenPLC. He bypassed the authentication using a default account and sent the crafted HTTP packets to the code box on the Hardware tab of the OpenPLC Runtime. Through the experiment, he could achieve control of the machine. The report revealed that the OpenPLC has a severe vulnerability in its code box and Webserver. After the report, the founder of the project responded by closing the problematic function and replacing it with Python SubModule (PSM) instead.

# 3 Vulnerabilities

We conducted an extensive examination of the OpenPLC project, with a specific focus on vulnerabilities within the Runtime application. After a comprehensive analysis, we have concluded that the project is susceptible to significant security threats. It exhibits deficiencies in key security aspects, particularly access control and encryption. Moreover, the application has neither intrusion detection nor mitigation mechanisms to resist cyber threats.

## 3.1 Opened Access to the OpenPLC project folder

Above all, the project folder permits access from users who are not necessarily related to the project. While the official documentation primarily suggests installing the software without requiring root privileges, from a security perspective, this recommendation cannot be recommended despite its distinct advantage of enhancing software availability. The software continues to allow access to all users on the OpenPLC-installed machine even after installing the software with a root privilege. It exposes essential components within the project folder to anyone with access to the OpenPLC-installed machine. Due to this issue, the software is prone to unauthorized access and potential attempts to compromise its functionality. While this approach may strengthen the availability of the software, it also increases security risks associated with unauthorized access.

## 3.2 Absence of User Access Level Control

SQLite3[1], the library utilized for the construction of the database referred to as "*openplc.db*", has a notable advantage in that it is lightweight disk-based because it does not require a server process. Nevertheless, it has a limitation not to support the function to manage the

---
[1]https://docs.python.org/3/library/sqlite3.html

```
kimchaer@kimchaer-VirtualBox:~/Documents$ python3 dbreader.py

*******Users********

10: OpenPLC User | btucs | openplc@openplc.com | kimchaer | None

kimchaer@kimchaer-VirtualBox:~/Documents$ python3 dbreader.py

******Programs******
PID: 22 | TITLE: program_1    | FILE: 348049.st | UPLOADED: Wed Feb 15 11:35:57 2023
PID: 23 | TITLE: program_2    | FILE: 453094.st | UPLOADED: Wed Feb 15 11:36:25 2023
PID: 24 | TITLE: original     | FILE: 111227.st | UPLOADED: Wed Feb 15 11:30:05 2023
PID: 26 | TITLE: user_program | FILE: 898031.st | UPLOADED: Wed Jul  5 14:54:01 2023
```

Figure 3.1: Reading database [1]

users' access level. It is inherent to the architecture of the SQLite3 database, allowing anyone in the machine to access and read a database. Accordingly, the project cannot ensure the confidentiality of critical information and the integrity of the control logic program, posing significant security concerns. Fig 3.1 refers to how credentials are stored in the database. By conducting experiments involving modifications to program identifiers, upload dates, and the location of program copies, it is confirmed, that any individual could access, read, and modify the database. See Fig 3.2. This feature poses a substantial risk of being abused by an adversary to deceive the user after he attempted the attack .

## 3.3 Plain Database Storage Format

As discussed in Chapter 2, the database *openplc.db* is responsible for storing pivotal information related to the entire OpenPLC system. However, it was primarily designed without security concerns, such as storing all information within the database in plain text, resulting in multiple critical vulnerabilities. This information encompasses user credentials, program data, registered devices, and other essential records. The lack of confidentiality could pose a significant security risk if this vulnerability combines with a lack of access control for the database. In other words, if individuals are capable of crafting the script to read the database, they can easily extract the critical information stored for the OpenPLC even without requiring root permission.

In addition, the openplc.db lacks the capability to reflect the update of the program. Although the OpenPLC permits the update of the uploaded programs, it does not update the program record in the database, particularly the date of upload. It is difficult for users to recognize whether the program is updated or not. Consequently, it is not feasible to notice whether the attacker silently updates the program. If the user updates the program, the software merely overwrites the updated version on the copy.

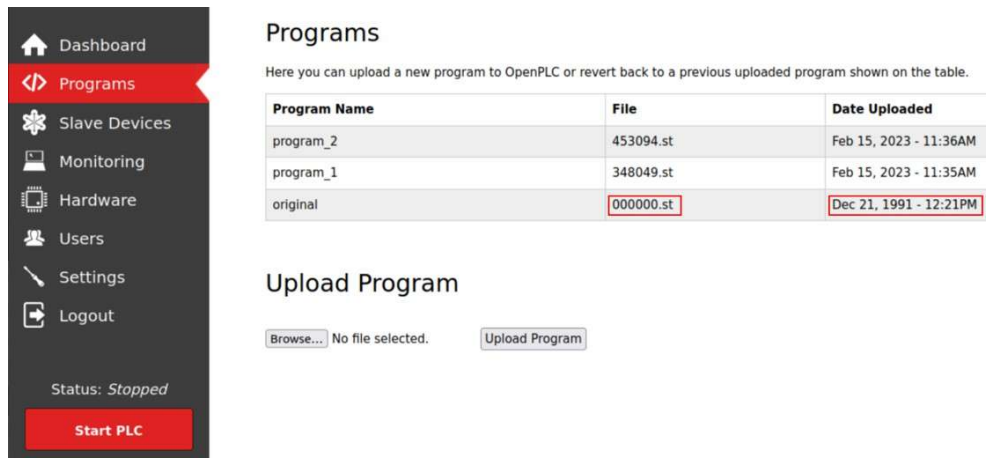

Figure 3.2: Manipulating database

## 3.4 Usage of vulnerable protocol

All of the security issues mentioned above pose critical threats, including the manipulation or addition of control logic programs, uploading an attacker's program, and interception of user credentials, among other risks. Nonetheless, the most severe disclosed security issue of the OpenPLC is that the software utilizes HTTP protocol to communicate with the operator.

HTTP was developed by Tim Berners-Lee in 1989 to transfer data. It became a cornerstone of data communication used for the World Wide Web. Although it was dominant in the past, more than 80% of websites utilize Hypertext Transfer Protocol Secure (HTTPS), the more secure version of HTTP nowadays. One of the significant shortcomings of the HTTP protocol is that it transfers all data in plain, unencrypted form, constituting a critical security concern. It is the most explicit security issue since anyone can observe whatever is transmitted between the server (OpenPLC) and the client (user). With the help of the Wireshark tool[2], an adversary can effortlessly obtain user credentials as well as files uploaded in ST format.

---
[2]https://www.wireshark.org/

Figure 3.3: Default credentials exposed in the HTTP packet [1]

Moreover, HTTP allows the repeated transmission of old packets because of its architectural feature that it is a stateless protocol. This structural characteristic implies that the HTTP server only keeps minimal information about its client. Accordingly, it is not feasible to determine whether the received request is legitimate or repeated from a different client, potentially an attacker. The transmitted packets lack mechanisms to defend against replay attacks, such as nonce, packet sequence, encryption, and timestamp. The weak nature of the HTTP protocol severely compromises the confidentiality and integrity of sending information and authentication of users. It is the most explicit security issue since anyone can observe the packet stream between the OpenPLC and the user.

The OpenPLC attempted to mitigate security concerns, such as malicious login attempts with stolen credentials, by expiring the session every 5 minutes. Nevertheless, the practicality of this approach is not as elevated as expected because of the unique feature of the HTTP protocol. By replicating all contents from the HTTP header that was sent previously, an attacker can achieve the control of OpenPLC.

## 3.5 A Risk of Remote Injection

According to [23], it is demonstrated that an adversary had the capability to manipulate the code box within the Hardware tab and establish a connection between the attacker's machine and the machine being installed OpenPLC. Although the founder of the project replaced the code box with PSM, we discovered that the possibility of being abused still exists. Specifically, if the artificial HTTP request contains Python code that corrupts files in the Webserver, it may compromise the integrity of the files. In addition, users will not be able to detect those activities.

As noted in Chapter 2, the Webserver plays a fundamental role in storing all scripts that serve the leading functions of the software. In other words, the script for controlling PSM is also stored in the Webserver. Assuming that an adversary possesses comprehensive knowledge of the structure of the Webserver and can execute replay attacks, using PSM includes the risk of remote exploitation by an attacker, introducing an additional vector to attack. We want to address this issue, though it has a few limitations.

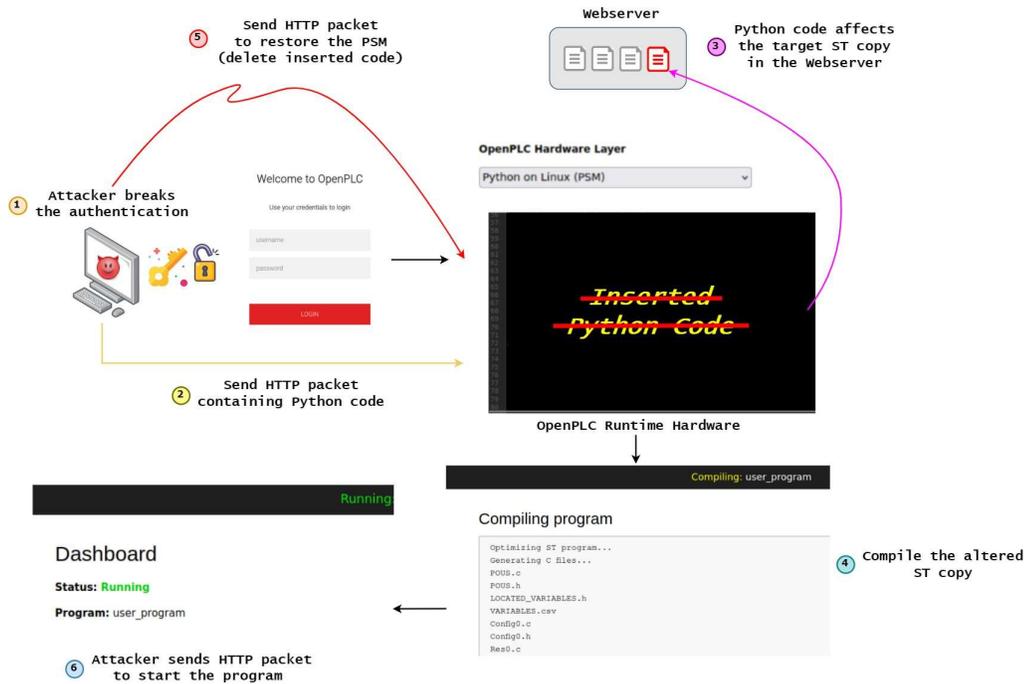

Figure 3.4: The overview of conducted experiment on PSM

In the conducted experiment, we transmitted the artificial HTTP request containing the Python script to modify the ST copy. Fig 3.4 shows the overview of the experiment exploiting the PSM box. Though the injection of the Python script left its trace in the Hardware tab, the user could detect the changes in the copy and kept running the program. Hence, we sent another HTTP request to restore the PSM. After transmission, the attack was still valid because restoring the PSM did not affect the ST copy. When the PSM saves the change on it, the Runtime application compiles the running program to reflect the change on the PSM. The software does not inform users about it. However, the attacker needs to be careful when he modifies the PSM because the Runtime log in the Dashboard prints the result of the execution of the PSM, like a terminal. Once the control logic is compiled, the software is prepared to execute the altered version. Although our attack scenario does not deploy the PSM for an attack, it still poses potential risks that enable adversaries to access files in the Webserver silently.

# 4  Our Attack

The previously presented works mainly focused on the falsification of messages between OpenPLC and SCADA software. Upon conducting an investigation into the operations of OpenPLC, however, we are of the opinion that their user interactions reveal significant vulnerabilities. They offer the possibility of being attacked by adversaries.

The ultimate goal of our approach is to damage the execution of the uploaded user program without being noticed by a user. Our approach introduces the exploitation of two vulnerabilities to implement the attack - the lack of access control on the Webserver and the weak nature of HTTP communication. Fig 4.1 shows a high-level overview of the implemented attack. To this end, the attack is launched in two phases: 1) Bypassing the authentication process and 2) Patching the application with the attacker's program. The adversary starts his attack from breaking the authentication and disguising himself as a legitimate user. After that, he identifies the target to start injection attack. The target is manipulated with help of identification process. After the adversary gives a damage to the target, he replays old legitimate packets to control the OpenPLC and execute the damaged target program. The most noticeable point of our approach is that it is silent and hard to be noticed by users. This attack manipulates the target without exposing any change on users' screen.

## 4.1  Attack scenario

1. **Bypassing the authentication process**
   First of all, the attacker needs a legitimate credential to bypass the authentication process because the OpenPLC Runtime application always requires authentication before it offers all types of services. He has two methods to break the authentication; monitoring the traffic by Wireshark tool or acquiring the credentials by accessing the *openplc.db* inappropriately.

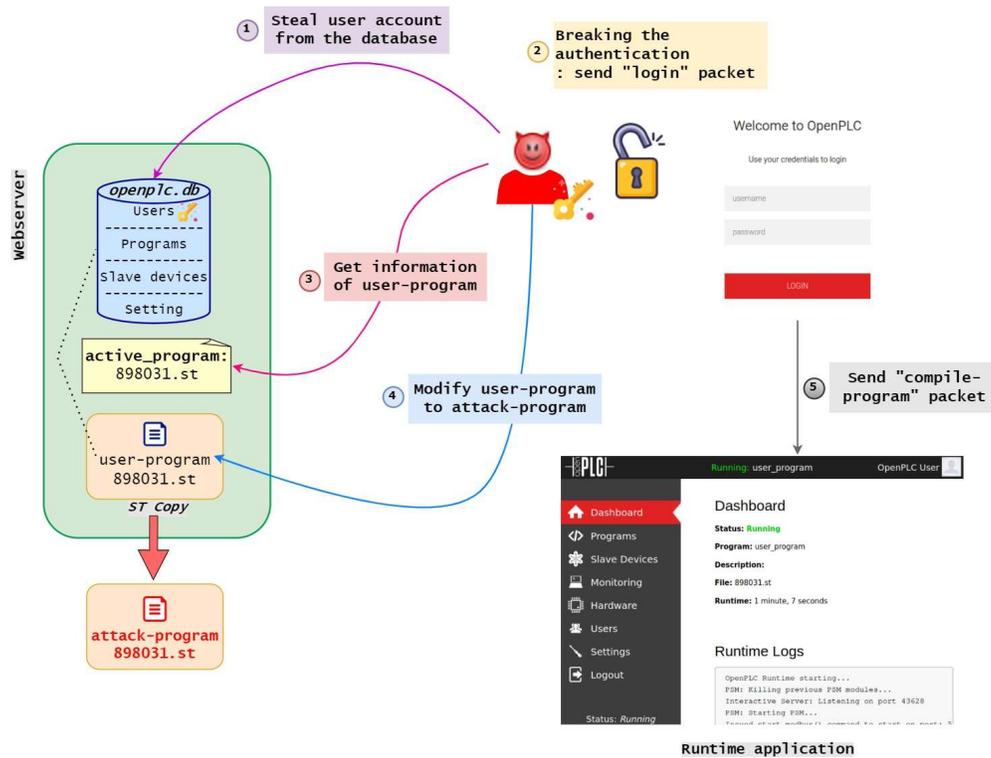

Figure 4.1: High-level overview of our attack approach [1]

a. *Replaying old packets*
When he monitors the network traffic between the operator and Runtime application, he may need a time to capture the exact packet transmitted for login from the user side (Fig 4.2). Once he discovered the credentials from the packet, the authentication can be done by replaying the old login packet including a legitimate credential. On the downside of this approach, the adversary has to wait until the packet for login is transmitted from the user side.

b. *Obtaining credentials by sneaking "openplc.db" file*
As mentioned, the main database, *openplc.db*, is freely opened to any users in the machine being installed the OpenPLC and has no mechanism to protect itself from the unauthorized access. Therefore, the attacker can access the openplc.db to get the user credentials instead of waiting until he captures the packet for login.

At first, he needs to find the location of the database since the OpenPLC can be installed anywhere user wants. The attacker can acquire the location of database with a simple script. Including uploaded programs' information, he can find out registered user accounts successfully from the database (Fig 4.3). With this approach, he enables taking the user credentials, as well as modifying or deleting the stored credentials. It results in legitimate users losing access to the service (Deny of Access).

Figure 4.2: Login packet

Figure 4.3: Information in "openplc.db" database [1]

2. **Patching the application**
   In order to significantly disrupt the execution of the OpenPLC, the target of attack should be the program running now, and the adversary needs to identify which program is executing currently before he starts to patch the target program. After that, he modifies the target and forces the Runtime application to run the attacker's program by transmitting artificially crafted packets.

a. *Identification of the target*

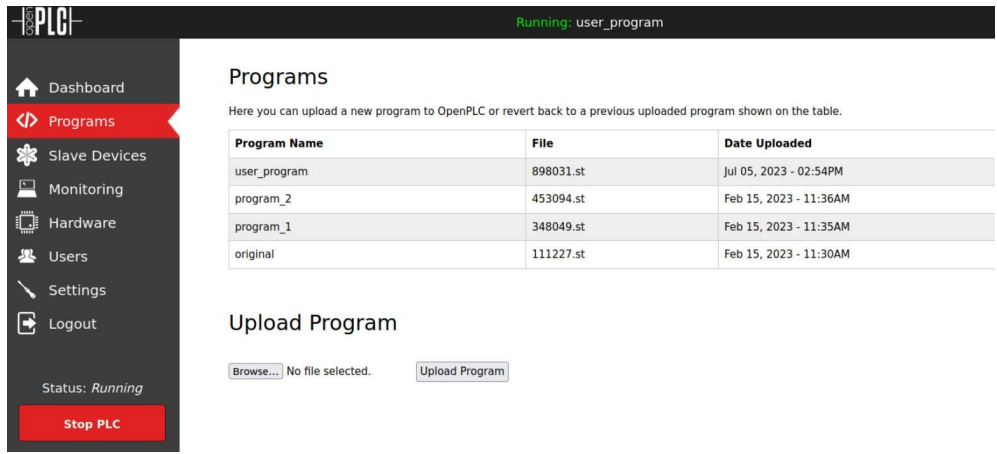

Figure 4.4: The program list [1]

Though the attacker could get the information of uploaded programs, the database does not present which program is running now. However, the information of currently running program is saved in the file titled 'active_program' that indicates which ST copy was the subject of the last compilation. When the OpenPLC compiles the program, it updates the title of copy in the 'active_program'. The 'active_program' is used as index to call the record of currently running program from the openplc.db when the OpenPLC starts its operation. Similar to other files in the Webserver, 'active_program' is accessible by any unauthorized user. Hence, the attacker can recognize which program is running now by accessing the index file. Furthermore, he can change or delete the title of copy in the index file and cause the operational error of the service (HTTP 500 status code).

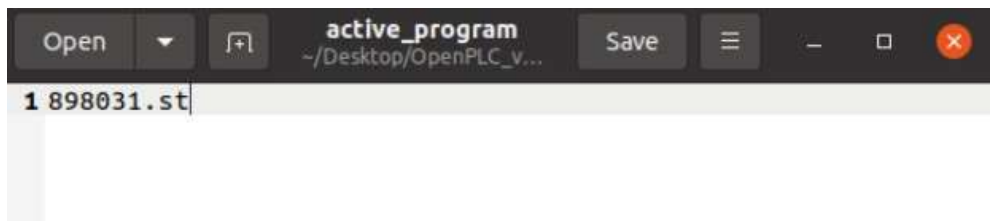

Figure 4.5: Reading the "active_program" [1]

Although checking 'active_program' is the simplest way to check the target, the attacker may choose another option to find out his target. If he monitors the packet stream between the operator and the Runtime application for enough time, he can discover the title of program which is running now from one of the "OK" packets (Fig 4.6). Once the adversary discovers the title of program, he can make the inference of the target by combining information of uploaded programs obtained from the database and title of program from the packet.

Figure 4.6: "OK" packets including the title of executing program [1]

Fig 4.6 is the response from the OpenPLC including the title of the currently running program. The Runtime application responds back with "OK" packet when it received the request to load the list of the program i.e., "program" packet. Loading the list of program is not the only method to discover the title of target. The target information is exposed in same manner when the operator gets the answer for the request to load the list of slave devices and users as well, i.e., "modbus" packet and "users" packet.

In our approach, the attacker can pick up the information of specific program which will be his target, with help of the 'active_program' (Fig 4.5). According to Fig 4.3, the *Prog_ID* of the target program is "26" and the ST copy of target is "898031.st". As mentioned in Chapter 2, the discovered Prog_ID will be used to load the program information when the adversary patches his program.

b. *Modifying the existing user-program*

   The attacker already has a full knowledge of the uploaded program by monitoring the packet traffic because the uploading program is completely exposed in plain text when the "upload-program" packet is transmitted. When the program is started, the Runtime application executes the compiled C code. Though the attacker may have the option to find out and update the C code directly, it is much simpler to modify ST format copy for him. The modification includes overwriting copy, inserting/deleting the instruction, changing the setpoint and the peripheral addressing, and replacing operators in equation.

c. *Patching the modified program*

Once the modification process is finished, the ST file needs to be uploaded on the software for the execution. At this point, the attacker has three options to proceed his attack. The first option is to make a new upload for his program. However, it remains a remarkable change that the operator may notice the abnormality. Uploading a new program remains a record in the openplc.db with a new unique identifiers, makes a new ST file copy with a new title in the Webserver, and causes packet traffic which may be considered an irregularity. See Fig 2.5. Furthermore, a new ST copy has a different title because it consists of randomly generated numbers. Even if we uploads the same program, the OpenPLC always creates a new record in openplc.db with different Prog_ID, title of ST copy, and the date of upload. Though the attacker can control the database, it requires more sophisticated effort to erase the footprints, such as fixing the Prog_ID, the date of upload, and the name of newly created ST file copy.

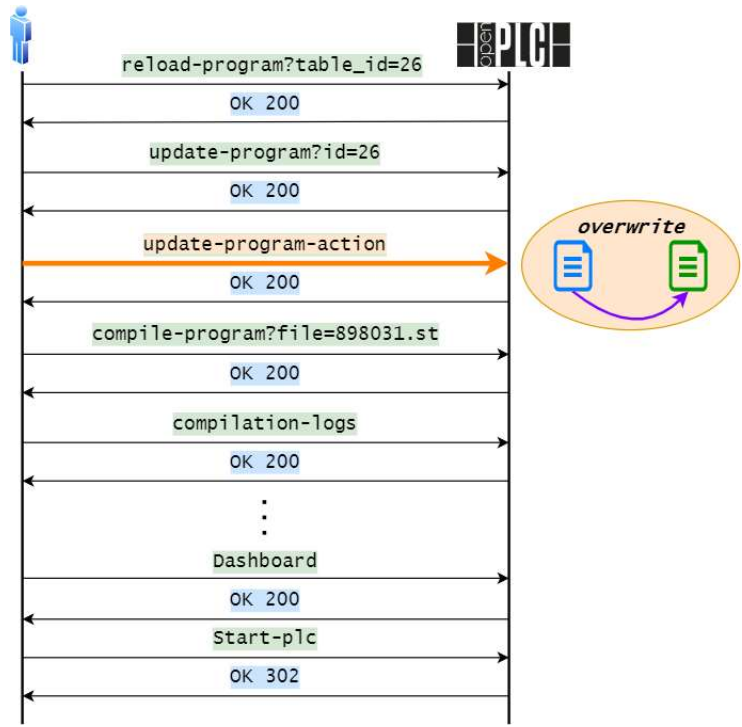

Figure 4.7: Updating process

Secondly, the attacker can update the existing program. After updating, the Prog_ID and the title of ST copy do not change, and the date of upload does not reflect the recent update as well since it is fixed on the uploaded date. However, we decided that it is not suitable to utilize for our approach though it does not remain any noticeable change on the OpenPLC. When the operator updates the existing program, he transmits the "reload-program?table_id=*Prog_ID*" to load the program information page and "update-program?id=*Prog_ID*" packets to load the page to select a new file for update. Then, the uploaded ST copy in the Webserver is overwritten by the selected file while the operator transmits the "update-program-action" packet. The compilation and execution process follow after the updating ST copy (Fig 4.7). The weak point of this approach is that, if the security manager monitors the packet traffic, he can have doubt for the packets sent for updating process and check whether it is suspicious activity or not.

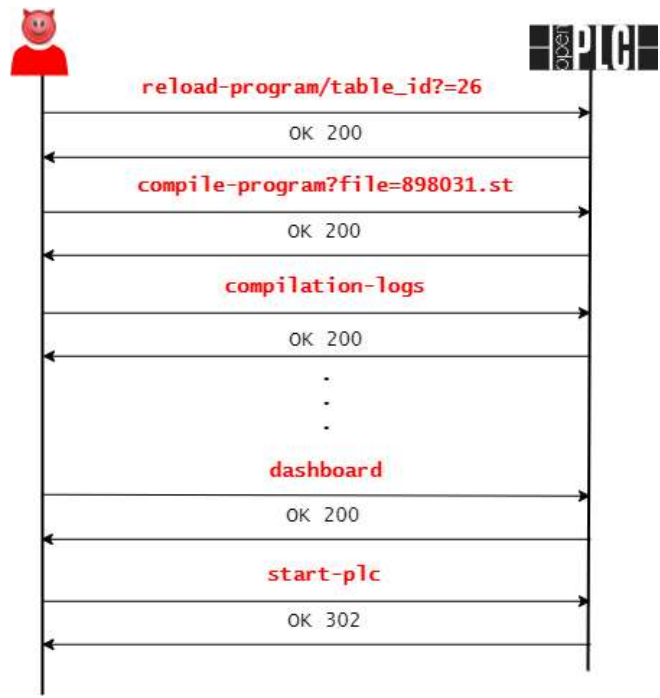

Figure 4.8: Patching the altered program [1]

Therefore, we decided to skip to re-transmit old packets generated from the uploading or updating process while implementing our attack and tried to establish the whole process simpler and stealthier. See Fig 4.8. In order to avoid detection of our injection attack, the adversary can simply compile the target file by transmitting the artificially crafted packets.

```
0220  5f 51 58 39 39 5f 31 09  76 61 72 54 79 70 65 3a   _QX99_1· varType:
0230  20 42 4f 4f 4c 0a 76 61  72 4e 61 6d 65 3a 20 5f    BOOL·va rName: _
0240  5f 51 58 30 5f 30 09 76  61 72 54 79 70 65 3a 20   _QX0_0·v arType: 
0250  42 4f 4f 4c 0a 43 6f 6d  70 69 6c 69 6e 67 20 6d   BOOL·Com piling m
0260  61 69 6e 20 70 72 6f 67  72 61 6d 2e 2e 2e 0a 43   ain prog ram...·C
0270  6f 6d 70 69 6c 61 74 69  6f 6e 20 66 69 6e 69 73   ompilati on finis
0280  68 65 64 20 73 75 63 63  65 73 73 66 75 6c 6c 79   hed succ essfully
0290  21 0a                                              !·
```

Figure 4.9: Last "OK" message during compilation process

Our strategy involves the attacker modifying the current ST copy that has been uploaded by the authorized user. Meaning that, the target program already has its identifier, the fixed date of upload, and the randomly generated title of the ST copy in the openplc.db. Compiling the currently running file and starting the program is all the attacker needs to do to execute the modified copy. Throughout the compilation process, the software consistently responds to "compilation-logs" packets to provide continuous updates. For every "compilation-logs" packet, the OpenPLC responds with an "OK" packet, and this "OK" packet includes messages generated while compiling. The last "OK" packet contains the message "compilation finished successfully". See Fig 4.9.

After compiling process, the software gets the C file of a modified ST file for its execution. Then, the adversary transmits the "start-plc" packet to execute the changed program. This approach allows the attacker to carry out a stealthy attack simplifying their efforts because the adversary utilizes the existing record in the database rather than generating a new record and leaving his traces. In addition, the patching process that our approach follows is the exact same method when the operator launches the program in ordinary cases. Our approach can be dismissed as an ordinary activity, even if the security manager monitors the packet traffic, as it appears to be a regular activity.

# 5  Security Countermeasures

In order to secure the OpenPLC project against all possibilities presented above, we suggest several methods in this section. The suggested security resolutions followed the ICS security guide [9] and are integrated in the presented new version – OpenPLC Aqua [4]. All information is exposed in plain text, and this fact gives a big piece of hint to attackers. Hence, the encryption should be entailed to keep the confidentiality while the software is executed. The encryption process is implemented not only in the openplc.db but also in the packet stream to conceal user credentials transmitting to the OpenPLC. Furthermore, OpenPLC has no access control for the Webserver. So, OpenPLC Aqua suggests strict access control during the operation. Lastly, we present a whitelisting function as a proactive method to block malicious activities and check the integrity and authenticity of newly uploaded files.

## 5.1  The AES algorithm

AES is a symmetric encryption algorithm, also known as Rijndael, established by the U.S. National Institute of Standards and Technology in 2001 [24]. It is selected to replace a traditional Data Encryption Standard (DES) algorithm that was used before and is a block-based cipher. The block-based cipher separates the plain text into several chunks at the first step of encryption, and the AES algorithm uses 128-bit message blocks. The algorithm allows three different lengths of key; 128, 192, and 256 bits. It repeats encryption rounds if the key becomes much longer. It is considered that AES-128 is faster and resource-efficient, while AES-256 is sturdier against brute-force attacks. However, both algorithms offer significant advantages in terms of security against modern threats.

Once the plaintext is divided into 128-byte blocks, the State Matrix is used to represent input and output as byte entries. The cipher is generated after four processing steps; SubBytes, ShiftRows, MixColumns, and AddRound. Initially, the round keys with 128-bit lengths are created from the key using AES key schedule. The input matrix is paired with the round key through a bitwise xor operation in every round. Then, each byte is converted using the

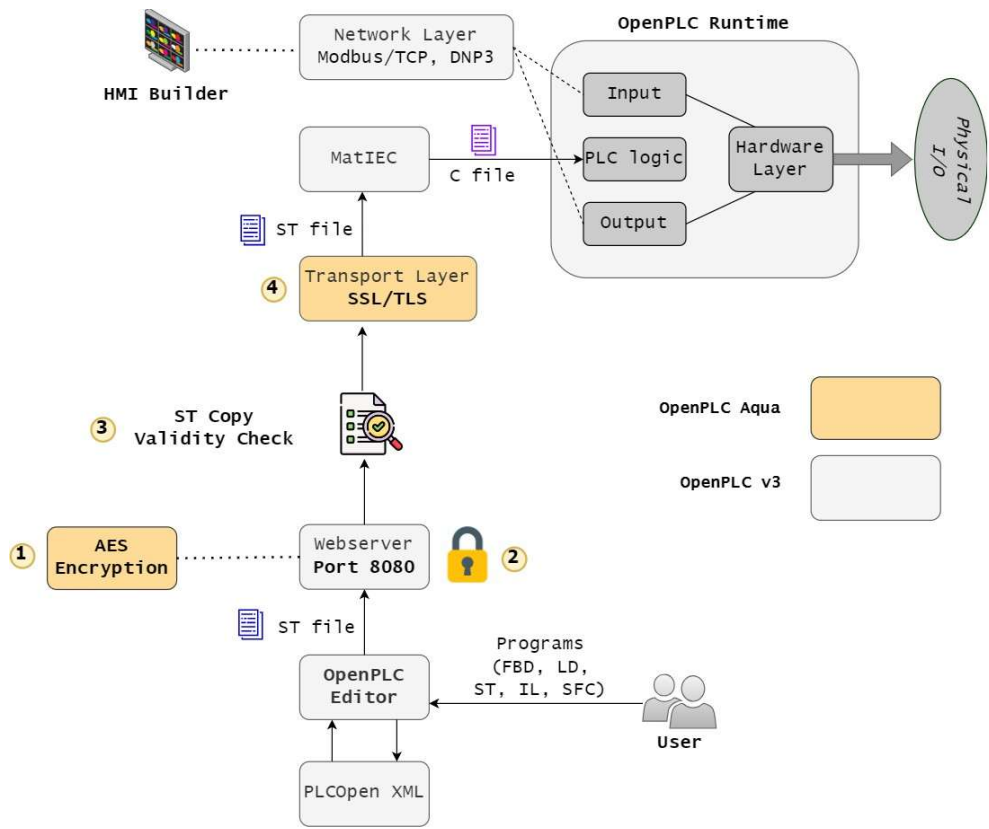

Figure 5.1: Architecture of OpenPLC Aqua [3]

S-box, an 8-bit substitution box, and the outputs of the S-box are shifted cyclically. Lastly, each column of the state is multiplied with a fixed matrix. The MixColumns step is skipped in the last round of encryption.

Whereas 3DES will not be allowed after 2023 due to its critical weaknesses, there is no practical attack against AES algorithms yet. Most presented attacks could not give a satisfactory performance to break AES until now.

## 5.2 Safe store of User credentials

The most criticized security concerns of the OpenPLC project are confidentiality, integrity, and authentication. Despite the requirement for user authentication for every action, the software remains vulnerable to bypass attacks. Thus, the security improvements for OpenPLC must include those three and not harm or modify the existing main features of the software.

Confidentiality ensures that critical information remains secure and inaccessible to any potential adversaries. Previously, the founder of the OpenPLC project used the AES algorithm to establish a secure SCADA communication in [2], and we considered it a great solution to protect user credentials from bypass attacks Fig 5.2.

There are several modes in AES encryption, and AES Ciphertext Block Chaining (AES-CBC) mode is chosen for the enhanced version because it has proven its stability for a long time. In the CBC mode, the message is separated into data blocks, and each block is encrypted with a ciphertext generated from the previous round. The ciphertext used to encrypt the first block is replaced with an Initial Vector (IV) in the initial round.

AES Galois/Counter Mode (AES-GCM) was another possible option offering a significant advantage of reducing overhead. However, we considered that the computational cost for encrypting user credentials would be almost the same between utilizing CBC- and GCM mode because the encryption process is only used to encrypt user credentials. Similarly, AES-GCM is a 128-bit block cipher providing data integrity and confidentiality as well [25]. While AES-CBC does not offer parallel processing, AES-GCM allows parallel processing. Therefore, it produces less overhead than AES-CBC. As input, it takes Key (K), Plaintext (P), and Associated Data (AD), and it results in Ciphertext (C), authentication Tag (T), and unencrypted AD. By checking T, a receiver can ensure that C and AD are not falsified. It utilizes AES-128 counter mode for encryption and arithmetic in the Galois Field to compute the T.

In our suggestion, it employs symmetric encryption, the AES-CBC algorithm, to support encryption when user credentials are stored in the database. While it is possible for attackers to obtain a username or password, it requires significant time and effort for them to find out a password in plain text.

OpenPLC Aqua uses only one pair of keys and IV for encryption, which means that the encrypted default account will have the same cipher strings because it has the same username and password. Despite considering the option to create another pair of key and IV, we ultimately decided against creating an additional set of encryption materials due to their memory consumption and perceived lack of necessity. The management of cryptographic materials was considered as well. Therefore, the software produces a single pair of a key and an IV during the installation. Both are 16 bytes in length and are generated randomly. In order to create different ciphers from the identical string with a single pair of a key and IV, our software employs a unique technique. When the software encrypts passwords, the key and IV are used for their original purpose when they encrypt passwords. Then, OpenPLC Aqua exchanges the location of key and IV to encrypt usernames. In other words, the software treats the IV as a key and the key as an IV. It is possible because they both are randomly generated bytes with equal length. As a result, it produces distinct ciphers from the identical string without additional encrypting materials and errors.

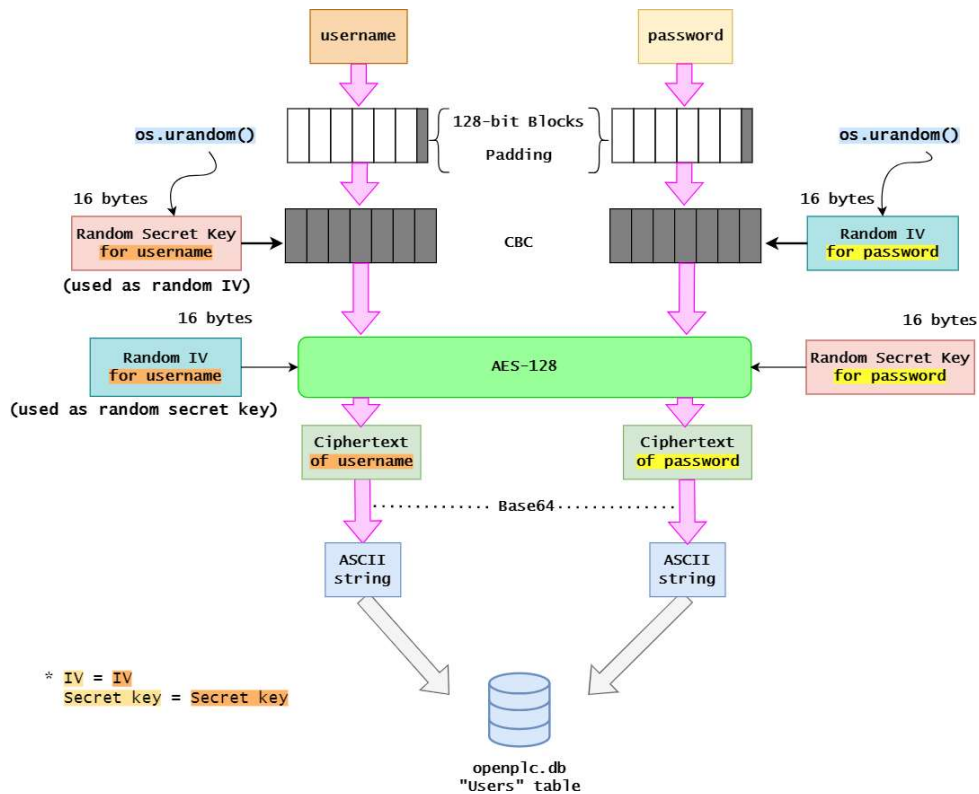

Figure 5.2: AES encryption in OpenPLC Aqua [4]

The result of AES-128 encryption consists of bytes which are technically not able to be stored in database. Thus, the ciphertext is encoded by Base64 algorithm and it is converted into ASCII letters. The Base64 algorithm is used to encode binary text to ASCII letters [26]. The most common usage of it is to encode an image.

The PLCs in the real world perform much more complex methodologies for their security. Not only for the management of user credentials, but also for detecting abnormal activities.

They provides functions such as rotating password or login with certificate to keep the user credentials securely. To find out anomaly, the commercial PLCs offer various forms of logs and this effort gives a dynamic view of PLCs' implementation. They record their outputs with peripheral addresses and provides various types of analysis using their accumulated log data. For instance, S7-1200 series support to store their activities with the address of memory and timestamp. Furthermore, the administrator can get a statistical analysis of their execution. To detect irregularity, our solution has the function to log the activities of the Runtime application as well.

## 5.3 SSL/TLS Handshake

The SSL/TLS Layer is used to establish a secure connection in numerous scenarios. The layer utilizes a handshake protocol to generate a confidential key shared between the client and server. This process requires trusted certificates from involved parties. If the server's certification solely suffices to establish a secure session, it is denoted as a one-way SSL/TLS connection. It is clarified as a two-way SSL/TLS connection when certificates from both client and server are mandated.

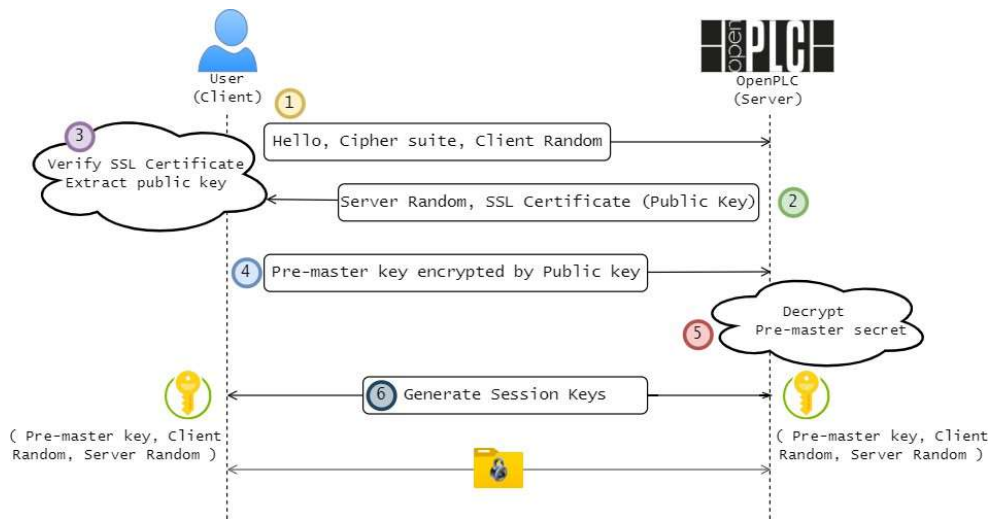

Figure 5.3: The one-way SSL/TLS Handshake [4]

When the client sends a "ClientHello" to inform that he wants to initiate the encrypted session, it contains the ClientRandom consisting of 32 bytes of random number along with cipher suites, and the version of SSL/TLS used. Then, the server responds "ServerHello" with its own ServerRandom, a distinct 32 bytes of random number, its chosen cipher suite, and its certificate. The client can extract the server's public key from the received certification and use it to encrypt a pre-master key. A pre-master key is a randomly generated 48-bit value to support deriving the keys for any cipher suites. Since SSL supports various types of encryption algorithms for key exchange, the pre-master key is to yield diverse lengths of keys. Then, the encrypted pre-master key is sent to the server to prove that the client can successfully derive the server's public key. Upon receiving the pre-master key, the server decrypts it using its private key, confirming the ability of the client to extract the accurate public key. With this verification, both parties can compute a matching session key utilizing their shared pre-master key, ClientRandom, and ServerRandom. The client transmits an encrypted message using the generated session key to validate their session key synchronization and conclude the handshake protocol. The server verifies the message by decrypting it with the server's session key and sends a signal that the server can also finish the handshake protocol. Subsequently, all communication between client and server is encrypted with the session key.

In the two-way SSL/TLS handshake protocol, both parties mutually verify each other's certificates. The verification of certifications consists of assessing digital signatures, the chain of certificates, the expiration and validity date, as well as its revocation status. The certificate chain is a list of certificates with the order, beginning from an SSL/TLS certificate to Certificate Authority (CA) certificates issued by publicly trusted authorities. It allows the receiver to verify the trustworthiness of the CA and the sender. Each certificate in the chain gets the signature from the entity of the following certification in the order. However, a self-signed certificate is deployed in some cases. Instead of being signed by CAs, a self-signed certificate gets a signature from developers or companies. As its root CA is empty, it is considered an unsafe certificate to use in ordinary. Hence, the web application alerts the client about this and lets them choose whether to proceed with the connection or not.

In the case of PLC products from SIEMENS, they utilize the TLS layer to establish a secure connection and encrypt the packet stream. Accordingly, their products issue certificates based on X.509 and keys for a secure TLS connection, and S7-1500 supports a self-signed certificate. By establishing a TLS layer over the HTTP, the security of user programs and devices is ensured against potential breaches and unauthorized access within industrial settings. The connection over TLS layer guarantees the confidentiality and integrity of data while the user and OpenPLC interact.

## 5.4  Secure Connection between User and OpenPLC

Adversaries are able to monitor lots of critical information in plain form, including user credentials and uploading user programs from the packet stream in the original OpenPLC. It is because that OpenPLC does not implement an SSL/TLS connection.

In our new version, OpenPLC Aqua, the HTTPS protocol is deployed for communication between the Runtime application and users. Accordingly, a pair of self-signed certificate and key is generated to establish a secure connection. The certificate is created while installing the software. After adopting an SSL/TLS connection, an entire packet stream is encrypted and prevented from sniffing attacks.

Table 5.1: HTTP vs. HTTPS [5]

| Features | HTTP | HTTPS |
|---|---|---|
| Use of SSL certificate | X | ✓ |
| Encryption | X | ✓ |
| Server Authentication | X | ✓ |
| Data Integrity | X | ✓ |
| Clear data transmission | X | ✓ |
| Performance Speed | ✓ | ✓ |

HTTPS has several advantages in the view of security. At first, it encrypts the transmitting data based on asymmetric encryption. Fig 5.3 shows how the data is encrypted using asymmetric encryption. Secondly, it verifies the authentication of the server. Having a private key means that it is an authenticated entity to communicate. It is verified by checking the validation of certificates. In addition, HTTPS ensures the integrity of data between the server and the client. Due to the features of HTTPS, the critical information in the HTTP packet's header and the body of the HTTP POST packet are encrypted. The data transmission becomes much clearer by using HTTPS, allowing entities to clarify the source of visitors. In the HTTP stream, the packet only shows the origin passed just before the arrival, e.g., Google Ads. In contrast, the HTTPS packet clearly denotes all sources it passed through before reaching the destination. In the past, deploying HTTPS could cause a delay when it initiates communication. Nonetheless, on average, Google processes over 40,000 queries per second through HTTPS these days. This fact indicates that HTTP and HTTPS have very minor differences in their speed for performance.

## 5.5 Access Control on Webserver

Since the prominent vulnerability that our presented attack exploited was free access to the Webserver, the access control on the Webserver has the priority to construct the protection of the software. In the real world, the most notable security solution for commercial PLCs is the access level restriction. Hence, we are of the opinion that the integration of an access control mechanism into the software can enhance the experiential quality for individuals desiring to acquire insights into the working process of PLCs.

As mentioned, the recent installation guide does not specify the recommendation to install the software with root permission, which causes severe vulnerabilities. Though it can increase the availability of the software, it is not helpful to improve the overall security level of the project. Accordingly, OpenPLC Aqua recommends users to install the software with a root privilege. It can offer a basic protection for the software against the access from unrelated users.

Whereas commercial PLCs configure distinct access levels for each user, the current OpenPLC does not allow users to designate access privileges. It is mostly because of the architectural problem of the database. Hence, we implemented the strong limitation of access on elements in the Webserver by executing the script to allow very limited access at the last step of installation. The strict restriction of access is implemented to the executable scripts in the Webserver that perform the primary functions of the Runtime application, as well as the database and list of registered IP addresses. The explication of the list of IP addresses is reserved for the following section.

## 5.6 Whitelisting

The above suggestions primarily focus on identifying and detecting suspicious activities. However, they do not have the capability to stop ongoing attacks. In addition, if an adversary obtains legitimate user credentials from a different source and abuses them for his successful login, OpenPLC demands an additional proactive measure to prevent their attack.

A concept of whitelisting is controlling the access very strictly by allowing only registered and legitimate entities with a policy that basically blocks all other attempts. For OpenPLC Aqua, a list of entities is generated during installation including a default pair of IP addresses and usernames. The default account for the first login is registered with the IP address of localhost (127.0.0.1). Upon enrolling a new user account, the software adds the username and current IP address of the current machine to the list automatically. Therefore, the login is allowed only if the entered username and current IP address are matched to the registered pair in the list.

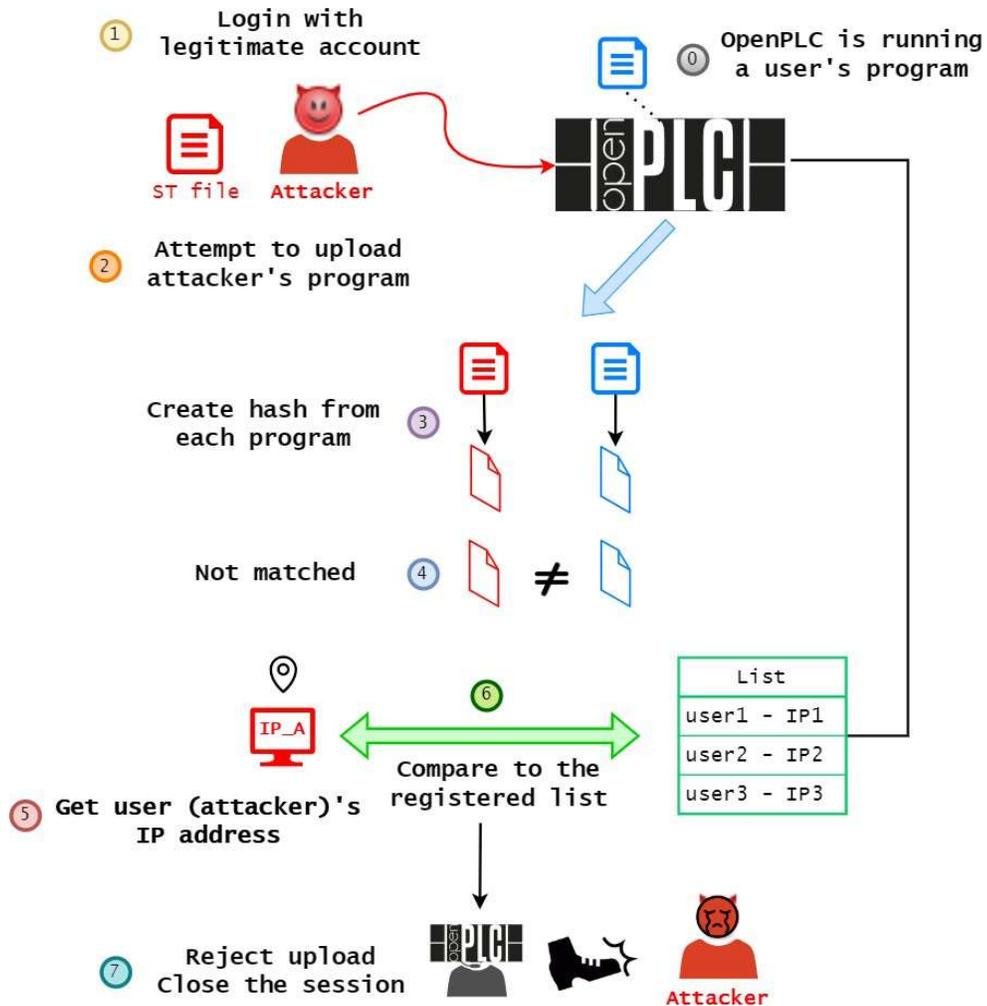

Figure 5.4: Whitelist function [4]

The assumed attacker may have a chance to obtain registered user credentials through certain routes. However, he is not aware of the content of the running program due to the encrypted SSL communication over HTTP. The final goal of the adversary is to access the OpenPLC and upload his malfunctioning program. In order to block the vicious attempt to upload a detrimental program, the Runtime application compares the file contents of the currently executing program with the newly uploaded program. From each file, the software generates Message-Digest Algorithm 5 (MD5) hashes and compares those two.

MD5 is a widely used message-digest algorithm that generates a 128-bit of hash [27]. In the past, MD5 was used for cryptographic purposes, but it had vulnerabilities that were not suitable for those purposes. Hence, it is deployed as a non-cryptographic function, such as a function to verify data integrity.

When the hashes are not matched, the bytes-by-bytes comparison is implemented while the "upload-in-action" packet is transmitted. If the software cannot find any difference between the two files, the upload process is proceeded without checking the current IP address of the user. When one of the comparison consequences results that the files are not identical, the OpenPLC inspects the current IP address of the user. The attempt to upload is frustrating if the pair of username and IP address that is currently active in the session does not match the enrolled list. The recent session is automatically closed right after the upload fails because the software doubts the user because of the user's activity. Finally, the software kicks out the suspicious user.

## 5.7 Removing the deleted copy

In the current version, when the operator requests to remove the program from the list, the program record is eliminated from the openplc.db. However, the ST copy of the deleted program remains in the Webserver forever and is irremovable. It may have the possibility to cause a potential security risk for adversaries to exploit remaining copies. Hence, we are of the opinion that ST copies should be removed as well. OpenPLC Aqua executes a functional script to remove the matching ST copy when an operator requests to delete the uploaded program from the list. Nevertheless, the compiled C file still exists in the Webserver and is ready to be executed on the Dashboard even after the program is eliminated from the database.

In addition, the software can cause an operational error at the very first step of its launch when the ST copy specified in the index file does not exist in the Websever, which means the currently executing program. It is because of the original feature of software that indexes the file stored in the 'active_program' when it launches. In order to solve the issue, a fake ST program is deployed temporarily. During the installation of OpenPLC Aqua, the default blank program is copied and named 'temporal_program'. It is utilized to replace the title of a currently executing program, and the software can prevent the execution of the eliminated program by replacing the title presented on the Dashboard. The copied program is configured not to be accessible by anyone except for root. Since the sole purpose of the bogus program is to ensure the successful launch of the Runtime application, the fake program is not allowed to be read, modified, or executed by any other users.

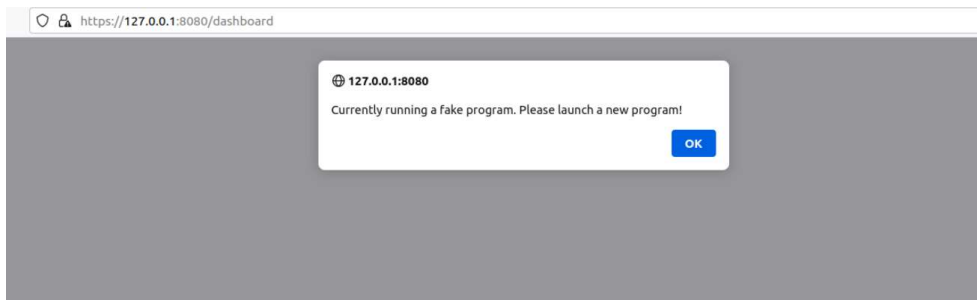

Figure 5.5: Alarm after the compiled program is eliminated

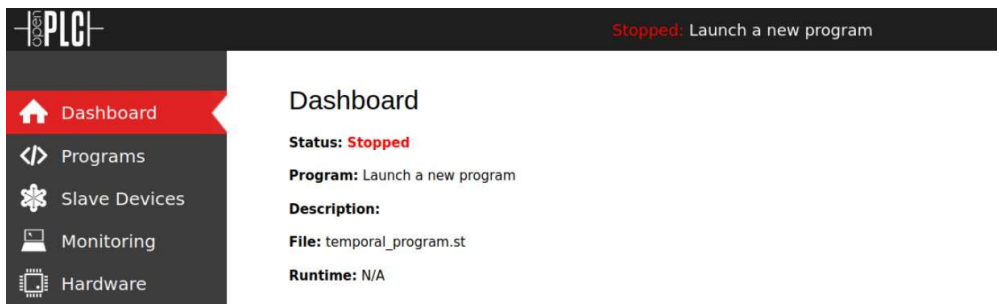

Figure 5.6: Dashboard when the compiled program is removed from the list

If the currently running program is eliminated from the list, the fake program is displayed on the Dashboard instead of it to prevent the execution of the deleted program. Moreover, the application sends alerts to users to compile the other programs. See Fig 5.5. Fig 5.6 shows the Dashboard when the compiled file is deleted. Therefore, it should deter compiling or executing the program when the fake program is on the Dashboard. OpenPLC Aqua configures that if the user attempts to run the bogus program, the software considers it suspicious activity and kicks out the current user, making the attacker hard to attack. It is designed to discourage the attempt to abuse a fake program and block the additional potential risk caused by having it.

# 6 Experimental Results and Discussions

In order to evaluate the performance of security features in OpenPLC Aqua and to compare the results, we conducted five distinct attack scenarios on the newly developed software and three different OpenPLC software variants: OpenPLCV3, OpenPLC Neo, and OpenPLC61850.

## 6.1 Experiments of various attacks against different OpenPLC versions

Table 6.1: The experiment result of various attacks against different OpenPLC versions [3]

| Attack scenario | v3 | Neo | 61850 | Aqua |
|---|---|---|---|---|
| Authentication Attack | ✓ | - | ✓ | - |
| Man-in-the-Middle Attack | ✓ | - | ✓ | - |
| Control Logic Injection Attack | ✓ | ✓ | ✓ | - |
| Replay Attack | ✓ | - | ✓ | - |
| Access Attack | ✓ | ✓ | ✓ | - |

Our experimental results are listed in Table 6.1. By adopting countermeasures to the project, OpenPLC Aqua evidently shows an improved performance for security compared to the original version. It neutralizes attempted attack scenarios, including the introduced control logic attack. At the same time, the software does not provide any hints for unauthorized users about critical information, such as registered user accounts, or control logic programs. It is feasible because the latest software implemented strict access controls on the user credentials within project folders, including *openplc.db* and Webserver. In order to launch attacks successfully, it is difficult for adversaries to find out other methods except for acquiring root permission. Even assuming he had access to the database coincidentally, he would only encounter encrypted usernames and passwords. Hence, decrypting processes for discovered ciphers is essential for him, and it requires additional efforts without the knowledge of encryption keys. The strict key management for encryption and rigid access control can frustrate attempts to take over legitimate accounts, adding a layer of security.

In contrast to the original OpenPLC, an attacker is not able to obtain any meaningful information about usernames and passwords from the packet traffic between the new software and legitimate users since the OpenPLC Aqua deploys the encrypted connection over a secure channel. It ensures both confidentiality and integrity of communicating data.

Moreover, the software offers a notable whitelisting function that prevents uploading attempts from untrustworthy users. If the attacker attempted to upload his ST program from the attacker's machine with a registered account, the software immediately dismisses the established connection between the attacker and the OpenPLC Runtime after verifying the validity of the ST file and user identity.

# 7 Conclusion

This thesis presented the existing security issues in the OpenPLC project, possible attacks exploiting them, and suggested the integrated resolutions, OpenPLC Aqua, to close the vulnerabilities. After conducting extensive research and experiments, we have confirmed that the OpenPLC remains vulnerable. The experiments showed that the adversary could manipulate the uploaded user program by breaking the authentication. The implemented attack included exploiting a lack of access control and the weak nature of HTTP communication to achieve the user credentials and retransmit old packets to control the software. The OpenPLC Aqua enhanced the level of security and stability of OpenPLC by integrating advanced security features. It offers encryption, access control, secured communication channels, and a safelist to prevent possible attacks. These measures ensure not only availability but also confidentiality and integrity of data. Though our solutions protect the communication between the user and software, it is not yet compatible with legacy SCADA software. We are of the opinion that our latest software can integrate the local gateway for SCADA software to enable the decryption of packets sent from the OpenPLC Aqua in the future. This thesis includes one publication that was published at the IECON 2023 conference and presented at the IACS WS'23 workshop.

# 8 Acronyms

**OpenPLC** Open-source Programmable Logic Controller
**PLC** Programmable Logic Controller
**ICS** Industrial Control System
**SCADA** Supervisory Control and Data Acquisition
**AES** Advanced Encryption Standard
**ASCII** American Standard Code for Information Interchange
**SSL/TLS** Secure Sockets Layer/Transport Layer Security
**DNS** Domain Name System
**DNP3** Distributed Network Protocol
**IEC** International Electrotechnical Commission
**FBD** Function Block Diagram
**LD** Ladder Diagram
**ST** Structured Text
**IL** Instruction List
**SFC** Sequential Function Chart
**XML** Extensible Markup Language
**HMI** Human Machine Interface
**TCP** Transmission Control Protocol
**HTTP** Hypertext Transfer Protocol
**SHA-256** Secure Hash Algorithm
**IPS** Intrusion Prevention System
**MMS** Manufacturing Message Specification
**FDI** False Data Injection
**FCI** False Command Injection
**MitM** Man-in-the-Middle
**ARP** Address Resolution Protocol
**PSM** Python SubModule
**HTTPS** Hypertext Transfer Protocol Secure
**DES** Data Encryption Standard
**AES-CBC** AES Ciphertext Block Chaining
**IV** Initial Vector
**AES-GCM** AES Galois/Counter Mode
**CA** Certificate Authority
**MD5** Message-Digest Algorithm 5

# 9 Appendix

For complete details on attacks and countermeasures, please refer to the videos ([28], [29]) and Git Respositories ([30], [31]).

# Bibliography

[1] Wael Alsabbagh, Chaerin Kim, and Peter Langendörfer. Good Night, and Good Luck: A Control Logic Injection Attack on OpenPLC. In *IECON 2023- 49th Annual Conference of the IEEE Industrial Electronics Society*, pages 1–8, 2023. doi: 10.1109/IECON51785. 2023.10312570. Singapore, Singapore.

[2] Thiago Alves, Thomas H. Morris, and Seong-Moo Yoo. Securing SCADA Applications Using OpenPLC With End-To-End Encryption. *Proceedings of the 3rd Annual Industrial Control System Security Workshop*, 2017.

[3] Wael Alsabbagh, Chaerin Kim, and Peter Langendörfer. Investigating the Security of OpenPLC: Vulnerabilities, Attacks, and Mitigation Solutions. In *IEEE Access*, 2023. doi: 10.13140/RG.2.2.34456.98566/2.

[4] Wael Alsabbagh, Chaerin Kim, and Peter Langendörfer. No Attacks Are Available: Securing the OpenPLC and Related Systems. In *IACS WS'23: 8th GI/ACM Workshop on Industrial Automation and Control Systems IACS*, 2023. doi: 10.13140/RG.2.2.24570. 47043. Berlin.

[5] Wikipedia Contributors. Hypertext Transfer Protocol secure - Wikipedia, the free encyclopedia. https://en.wikipedia.org/wiki/HTTPS, .

[6] Wikipedia Contributors. Stuxnet - Wikipedia, the free encyclopedia. https://en.wikipedia.org/wiki/Stuxnet, .

[7] BBC News. Hack attack causes 'massive damage' at steel works. https://www.bbc.com/news/technology-30575104.

[8] Wikipedia Contributors. Industroyer - Wikipedia, the free encyclopedia. https://en.wikipedia.org/wiki/Industroyer, .

[9] Omar EL Idrissi, Abdellatif Mezrioui, and Abdelhamid Belmekki. Cyber security challenges and issues of industrial control systems–some security recommendations. In *2019 IEEE International Smart Cities Conference (ISC2)*, pages 330–335, 2019. doi: 10.1109/ISC246665.2019.9071701.

[10] Francisco Rodríguez, José Luis Guzmán, María del Mar Castilla, Jorge Antonio Sánchez-Molina, and Manuel Berenguel. A proposal for teaching SCADA systems using Virtual Industrial Plants in Engineering Education. *IFAC-PapersOnLine*, 49(6): 138–143, 2016. ISSN 2405-8963. doi: https://doi.org/10.1016/j.ifacol.2016.07.167. URL https://www.sciencedirect.com/science/article/pii/S240589631630372X. 11th IFAC Symposium on Advances in Control Education ACE 2016.

[11] Thiago Alves, Mario Buratto, Flavio Mauricio de Souza, and Thelma Virginia Rodrigues. OpenPLC: An open source alternative to automation. In *IEEE Global Humanitarian Technology Conference (GHTC 2014)*, pages 585–589, 2014. doi: 10.1109/GHTC.2014.6970342.

[12] Michael Tiegelkamp and Karl-Heinz John. *IEC 61131–3: Programming Industrial Automation Systems*. Springer, 1995.

[13] Wikipedia Contributors. Structured Text - Wikipedia, the free encyclopedia. https://en.wikipedia.org/wiki/Structured_text, .

[14] Mário de Sousa and Adriano Carvalho. An IEC 61131-3 compiler for the matPLC. In *EFTA 2003. 2003 IEEE Conference on Emerging Technologies and Factory Automation. Proceedings (Cat. No.03TH8696)*, volume 1, pages 485–490 vol.1, 2003. doi: 10.1109/ETFA.2003.1247746.

[15] Mário de Sousa. MatPLC-the truly open automation controller. In *IEEE 2002 28th Annual Conference of the Industrial Electronics Society. IECON 02*, volume 3, pages 2278–2283, 2002. doi: 10.1109/IECON.2002.1185327.

[16] Wikipedia Contributors. Modbus - Wikipedia, the free encyclopedia. https://en.wikipedia.org/wiki/Modbus, .

[17] Wikipedia Contributors. DNP3 - Wikipedia, the free encyclopedia. https://en.wikipedia.org/wiki/DNP3, .

[18] Thiago Alves, Rishabh Das, and Thomas Morris. Embedding Encryption and Machine Learning Intrusion Prevention Systems on Programmable Logic Controllers. *IEEE Embedded Systems Letters*, 10(3):99–102, 2018. doi: 10.1109/LES.2018.2823906.

[19] Muhammad M. Roomi, Wen Shei Ong, Daisuke Mashima, and Suhail Hussain. OpenPLC61850: An IEC 61850 compatible OpenPLC for Smart Grid Research. *SoftwareX*, 17, 2022. doi: 10.1016/j.softx.2021.100917.

[20] Wikipedia Contributors. Iec 61850 - wikipedia, the free encyclopedia. [https://en.wikipedia.org/wiki/IEC_61850](https://en.wikipedia.org/wiki/IEC_61850), .

[21] Muhammad M. Roomi, Wen Shei Ong, Suhail Hussain, and Daisuke Mashima. IEC 61850 Compatible OpenPLC for Cyber Attack Case Studies on Smart Substation Systems. *IEEE Access*, 10:9164–9173, 2022. doi: 10.1109/ACCESS.2022.3144027.

[22] Wael Alsabbagh, Samuel Amogbonjaye, Diego Urrego, and Peter Langendörfer. A Stealthy False Command Injection Attack on Modbus based SCADA Systems. In *2023 IEEE 20th Consumer Communications Networking Conference (CCNC)*, pages 1–9, 2023. doi: 10.1109/CCNC51644.2023.10059804.

[23] CVE-2021-31630. [https://cve.mitre.org/cgi-bin/cvename.cgi?name=CVE-2021-31630](https://cve.mitre.org/cgi-bin/cvename.cgi?name=CVE-2021-31630).

[24] Wikipedia Contributors. Advanced Encryption Standard - wikipedia, the free encyclopedia. [https://en.wikipedia.org/wiki/Advanced_Encryption_Standard](https://en.wikipedia.org/wiki/Advanced_Encryption_Standard), .

[25] Wikipedia Contributors. Galois/Counter Mode - Wikipedia, the free encyclopedia. [https://en.wikipedia.org/wiki/Galois/Counter_Mode](https://en.wikipedia.org/wiki/Galois/Counter_Mode), .

[26] Wikipedia Contributors. Base64 - Wikipedia, the free encyclopedia. [https://en.wikipedia.org/wiki/Base64](https://en.wikipedia.org/wiki/Base64), .

[27] Wikipedia Contributors. MD5 - Wikipedia, the free encyclopedia. [https://en.wikipedia.org/wiki/MD5](https://en.wikipedia.org/wiki/MD5), .

[28] [https://youtu.be/rEBeV982gWQ](https://youtu.be/rEBeV982gWQ), .

[29] [https://youtu.be/knVTQfUdNfU](https://youtu.be/knVTQfUdNfU), .

[30] https://github.com/rnrn0909/A-Control-Logic-Injection-Attack-on-OpenPLC.git, .

[31] https://github.com/rnrn0909/OpenPLC-Aqua.git, .